\documentstyle[psfig]{mn}
\newif\ifAMStwofonts

\def\HI{\hbox{H\,{\sc i}}}
\def\HII{\hbox{H\,{\sc ii}}}
\def\arcdeg{\hbox{$^\circ$}}

\def\kms{km~s$^{-1}$}
\def\aap{A\&A}
\def\mnras{MNRAS}
\def\apj{ApJ}

\def\aj{AJ}
\def\pasj{PASJ}
\def\pasp{PASP}
\def\pasa{PASA}

\def\nat{Nature}

\ifoldfss
  \ifCUPmtlplainloaded \else
    \NewTextAlphabet{textbfit} {cmbxti10} {}
    \NewTextAlphabet{textbfss} {cmssbx10} {}
    \NewMathAlphabet{mathbfit} {cmbxti10} {} 
    \NewMathAlphabet{mathbfss} {cmssbx10} {} 
  \fi
  \ifAMStwofonts
    \ifCUPmtlplainloaded \else
      \NewSymbolFont{upmath} {eurm10}
      \NewSymbolFont{AMSa} {msam10}
      \NewMathSymbol{\upi}     {0}{upmath}{19}
      \NewMathSymbol{\umu}     {0}{upmath}{16}
      \NewMathSymbol{\upartial}{0}{upmath}{40}
      \NewMathSymbol{\leqslant}{3}{AMSa}{36}
      \NewMathSymbol{\geqslant}{3}{AMSa}{3E}

    \fi
  \fi
\fi 

\ifnfssone
  \newmathalphabet{\mathit}
  \addtoversion{normal}{\mathit}{cmr}{m}{it}
  \addtoversion{bold}{\mathit}{cmr}{bx}{it}
  \newmathalphabet{\mathbfit} 
  \addtoversion{normal}{\mathbfit}{cmr}{bx}{it}
  \addtoversion{bold}{\mathbfit}{cmr}{bx}{it}
  \newmathalphabet{\mathbfss} 
  \addtoversion{normal}{\mathbfss}{cmss}{bx}{n}
  \addtoversion{bold}{\mathbfss}{cmss}{bx}{n}
  \ifAMStwofonts
    \ifCUPmtlplainloaded \else
      \UseAMStwoboldmath
      \makeatletter
      \new@mathgroup\upmath@group
      \define@mathgroup\mv@normal\upmath@group{eur}{m}{n}
      \define@mathgroup\mv@bold\upmath@group{eur}{b}{n}
      \edef\UPM{\hexnumber\upmath@group}
      \new@mathgroup\amsa@group
      \define@mathgroup\mv@normal\amsa@group{msa}{m}{n}
      \define@mathgroup\mv@bold\amsa@group{msa}{m}{n}
      \edef\AMSa{\hexnumber\amsa@group}
      \makeatother
      \mathchardef\upi="0\UPM19
      \mathchardef\umu="0\UPM16
      \mathchardef\upartial="0\UPM40
      \mathchardef\leqslant="3\AMSa36
      \mathchardef\geqslant="3\AMSa3E
    \fi
  \fi
\fi 

\ifnfsstwo
  \DeclareMathAlphabet{\mathbfit}{OT1}{cmr}{bx}{it}
  \SetMathAlphabet\mathbfit{bold}{OT1}{cmr}{bx}{it}
  \DeclareMathAlphabet{\mathbfss}{OT1}{cmss}{bx}{n}
  \SetMathAlphabet\mathbfss{bold}{OT1}{cmss}{bx}{n}
  \ifAMStwofonts
    \ifCUPmtlplainloaded \else
      \DeclareSymbolFont{UPM}{U}{eur}{m}{n}
      \SetSymbolFont{UPM}{bold}{U}{eur}{b}{n}
      \DeclareSymbolFont{AMSa}{U}{msa}{m}{n}
      \DeclareMathSymbol{\upi}{0}{UPM}{"19}
      \DeclareMathSymbol{\umu}{0}{UPM}{"16}
      \DeclareMathSymbol{\upartial}{0}{UPM}{"40}
      \DeclareMathSymbol{\leqslant}{3}{AMSa}{"36}
      \DeclareMathSymbol{\geqslant}{3}{AMSa}{"3E}
    \fi
  \fi
\fi 

\ifCUPmtlplainloaded \else
  \ifAMStwofonts \else 
    \def\upi{\pi}
    \def\umu{\mu}
    \def\upartial{\partial}
  \fi
\fi

\title[HI structure of the LMC]{A New Look at the Large-Scale \HI\ Structure of the LMC}
\author[L. Staveley-Smith et al.]
       {L.~Staveley-Smith,$^1$\thanks{email: Lister.Staveley-Smith@csiro.au} S.~Kim,$^2$ M.~R.~Calabretta,$^1$ R.~F.~Haynes,$^{1,3}$ 
\newauthor and M.~J.~Kesteven$^1$\\
$^1$Australia Telescope National Facility, CSIRO, P.O. Box 76, Epping,
NSW 1710, Australia\\
$^2$Harvard Smithsonian Center for Astrophysics, 60 Garden Street,
Cambridge, MA 02138, U.S.A\\
$^3$School of Mathematics and Physics, University of Tasmania,
   P.O. Box 252-37, Hobart, Tasmania 7001, Australia
}
\date{Accepted 2002 October 8.
      Received 2002 July 24}

\pagerange{\pageref{firstpage}--\pageref{lastpage} As accepted.}
\pubyear{2002}

\begin{document}

\maketitle

\label{firstpage}

\begin{abstract}
  We present a Parkes multibeam \HI\ survey of the Large Magellanic
  Cloud (LMC). This survey, which is sensitive to spatial structure in
  the range 200 pc $\sim$ 10 kpc, complements the Australia Telescope
  Compact survey, which is sensitive to structure in the range 15 pc
  $\sim$ 500 pc. With an rms column density sensitivity of $8\times
  10^{16}$ cm$^{-2}$ for narrow lines and $4\times 10^{17}$ cm$^{-2}$
  for typical linewidths of 40 \kms, emission is found to be extensive
  well beyond the main body of the LMC.  Arm-like features extend from
  the LMC to join the Magellanic Bridge and the Leading Arm, a forward
  counterpart to the Magellanic Stream. These features, whilst not as
  dramatic as those in the SMC, appear to have a common origin in the
  Galactic tidal field, in agreement with recent 2MASS and DENIS
  results for the stellar population. The diffuse gas which surrounds
  the LMC, particularly at pa's 90\degr\ to 330\degr, appears to be
  loosely associated with tidal features, but loosening by the ram
  pressure of tenuous Galactic halo gas against the outer parts of the
  LMC cannot be discounted.  High-velocity clouds, which lie between
  the Galaxy and the LMC in velocity and which appear in the UV
  spectra of some LMC stars, are found to be associated with the LMC
  if their heliocentric velocity exceeds about $+100$ \kms.  They are
  possibly the product of energetic outflows from the LMC disk.  The
  \HI\ mass of the LMC is found to be $(4.8\pm0.2)\times 10^8$
  M$_{\sun}$ (for an assumed distance of 50 kpc), substantially more
  than previous recent measurements.
\end{abstract}


\section{INTRODUCTION}

The LMC plays a key role in our understanding of diverse areas in
astronomy, including the extragalactic distance scale which uses the
LMC as a zero-point (Feast 1999, Gibson 1999), the formation of star
clusters (Johnson et al. 1999) and \HII\ regions (Oey 1996),
molecular cloud astrophysics (Johansson et al. 1994), and for
providing background stars with which to study possible microlenses in
the Galactic halo (Alcock et al. 2000). With this in mind, Kim et al.
(1998a) surveyed the LMC in \HI\ at high spatial resolution with the
Australia Telescope Compact Array (ATCA). This data has since served
to help study the interaction between star-forming regions and the
interstellar medium at small scales (e.g.  Kim et al. 1998b; Kim et
al.  1999; Points et al.  2000; Olsen, Kim \& Buss  2001). However its use
for large-scale studies (with notable exceptions, e.g.  the study of
the LMC disk and halo dynamics by Alves \& Nelson 2000) is largely
confined to morphological studies and comparisons (e.g. Wada, Spaans \& Kim
2000). The reason for this is that, being an interferometer, the ATCA
is insensitive to structure on an angular size scale larger than
$\lambda/B$, where the shortest baseline is $B=30$ m ($\sim20$ m when
the finite antenna size is taken into consideration and when the sky
is Nyquist-sampled by the antenna primary beams). This corresponds to
$\sim0\fdg5$ or about 0.4 kpc at the distance of the LMC (assumed here
to be 50 kpc).

The missing large-scale structure means that \HI\ column densities are
difficult to derive. For example, accurate optical
depths for X-ray photoelectric absorption cannot be obtained. It also
makes it difficult to compare the \HI\ structure of large shells against
competing models of their formation, e.g. stellar winds (Dopita, Mathewson \& Ford
1985), high-velocity cloud collisions (Braun 1996) and gamma-ray
bursts (Efremov, Ehlerov\'{a} \& Palou\u{s} 1999). Similarly, the
outer tidal structure of the LMC cannot be compared with the recent
stellar results from 2MASS and DENIS (van der Marel 2001).

Observations sensitive to large spatial scales may be taken from the
autocorrelations from the individual antennas of an interferometer.
But this data is often not calibrated in a suitable way. Moreover, for
a homogeneous interferometer such as the ATCA, there exists a serious
gap in the UV-plane between the auto and cross-correlation data. It is
usually more useful to collect data from a large single-dish antenna
such as the Parkes telescope\footnote{The Parkes telescope is part of
  the Australia Telescope which is funded by the Commonwealth of
  Australia for operation as a National Facility managed by CSIRO.}
which, with a diameter of 64 m, is much larger than the smallest ATCA
baseline. There are several excellent studies using Parkes of HI in
the LMC, however none is particularly suitable for the present
purpose. Needless to say, the older studies (McGee \& Milton 1966) are
not available digitally.  The observations of Luks \& Rohlfs (1992) do
not cover a sufficient spatial area (they cut off the LMC disk north
of Dec. $-66\arcdeg$) and are not Nyquist-sampled. The HIPASS
observations (Putman et al. 2002) have insufficient velocity resolution and,
since HIPASS was designed to detect compact extragalactic objects
(Barnes et al. 2001), also impose a high-pass filter on the sky.

The task of mapping a large area such as the LMC at the Nyquist rate
(5\farcm7 for Parkes at $\lambda21$ cm) is not straightforward as
$\sim 10^4$ pointings are required. Fortunately, the advent of a 21 cm
multibeam receiver at Parkes (Staveley-Smith et al. 1996) makes the
task easier, so observations were undertaken following the installation of
new narrow-band filters in late-1998. We describe the Parkes
observations in Section~\ref{s:obs} and discuss the general
morphology, tidal features and spatially integrated properties in
Section~\ref{s:results}. We also compare our results with previous
work. In Section~\ref{s:hvgas}, we discuss halo gas in the LMC,
particularly in the sightline from our Galaxy.  In
Section~\ref{s:holes}, we look again at the largest \HI\ holes and
associated high-velocity gas.  Finally, in Section~\ref{s:galaxy}, we
use the Galactic part of the velocity range to re-discuss foreground
absorption. The combination of the existing data with the ATCA data is
separately described in Kim et al.  (2002), and some early results
from the combined data set are discussed by Elmegreen, Kim \&
Staveley-Smith (2001) and Padoan et al. (2001).

\section{OBSERVATIONS}
\label{s:obs}

\begin{figure}
 \centerline{\psfig{figure=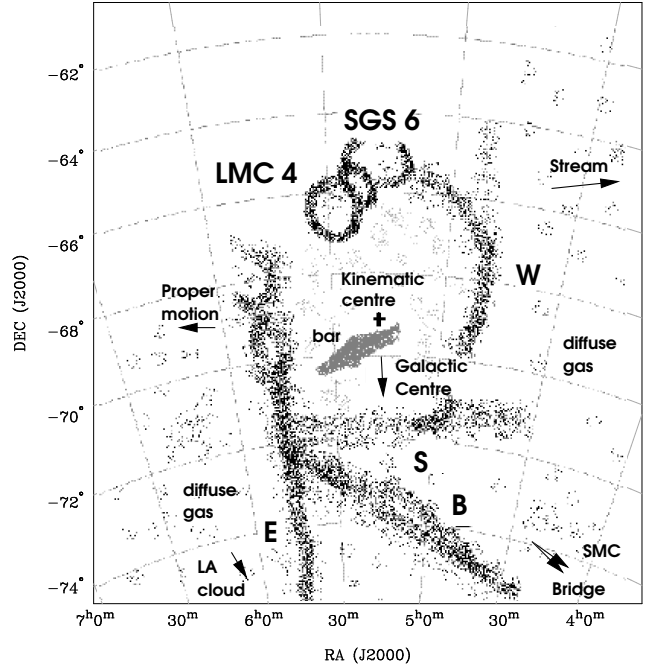,width=\columnwidth}}
 \caption{A schematic showing the main outer features of the LMC in \HI.
   The kinematic centre of Kim et al. (1998a) and the optical bar are
   included for reference. Arms B, E, S and W are the outer arms
   referred to in the text; LMC 4 (Meaburn 1980) and LMC SGS 6 (Kim et al.
   1999) are supergiant shells; diffuse gas appears in the south-east,
   south and west of the LMC. The directions to the Galactic Centre,
   the Magellanic Bridge, the SMC, the Leading Arm cloud at
   $\ell=292\fdg4$, $-30\fdg2$ and the Magellanic Stream (a vector at
   constant Magellanic latitude $-5\degr$, Putman et al. 2002) are shown. Also
   shown is the proper motion vector of Kroupa \& Bastian (1997).}

 \label{f:arms}
\end{figure}

Observations were taken with the inner seven beams of the Parkes 21 cm
multibeam receiver \cite{ss96} on 1998 December 13 to 17. The
telescope was scanned across the LMC in orthogonal great circles
aligned approximately east-west and north-south. The receiver was
continuously rotated such that the rotation angle was always at
$19\fdg1$ to the scan trajectory, the appropriate angle in a hexagonal
geometry for ensuring uniform spatial sampling of the sky. The area
covered was 13\degr\ by 14\degr\ in RA and Dec, respectively, and
centred on RA $05^{\rm h} 20^{\rm m}$, Dec. $-68\degr 44\arcmin$
(J2000). This corresponds to 8--9 disk scale lengths in
the optical or infrared (Bothun \& Thompson 1988, van der Marel 2001).
In a single scan, the spacing between adjacent tracks is 9\farcm5,
which is smaller than the mean FWHP beamwidth of 14\farcm1, but
greater than the Nyquist interval ($\lambda/2D$) of 5\farcm7.
Therefore, six scans were interleaved in each of the principal scan
directions, resulting in a final track spacing of 1\farcm6. In total,
$12\times 6$ RA scans and $11\times 6$ Dec. scans were made. Seven
scans were dropped or edited out due to drive problems, leaving a
total of 131 scans consisting of a total of 29 h of on-source
integration on each of seven beams. The average integration time per beam
area is 360 s (both polarisations).

The scan rate of the telescope was 1\fdg0 min$^{-1}$ and the
correlator was read every 5~s. Therefore, the beam was slightly
broadened in the scan direction to 14\farcm5 \cite{b01}.
After averaging orthogonal scans, the effective beamwidth reduces to
14\farcm3.  The central observing frequency was switched between
1417.5 and 1421.5 MHz, again every 5~s. This allowed the bandpass
shape to be calibrated without spending any time off-source. A
bandwidth of 8 MHz was used with 2048 spectral channels in each of two
orthogonal linear polarisations. \HI\ emission from the LMC (and the Galaxy)
appeared
within the band at both frequency settings. After bandpass
calibration, the data from both settings were shifted to a common
heliocentric reference frame. The velocity spacing of the
multibeam data is 0.82 \kms, but the final cube was Hanning-smoothed
to a resolution of 1.6 \kms.  The useful velocity range in the final
cube (i.e. after excluding frequency sidelobes of the LMC and the
Galaxy, and band-edge effects) is $-66$ to 430 \kms.

Bandpass calibration, velocity shifting and preliminary spectral
baseline fitting were all done using the {\sc aips++} {\sc LiveData} task. 
Baselines were adaptively fitted using polynomials of degree 8.
Subsequently, the data were convolved onto a grid of 4\arcmin\
pixels using a Gaussian kernel with a FWHP of 8\farcm0. This
broadens the effective, scan-broadened, beamwidth of the inner seven beams
from 14\farcm3 to about 16\farcm4. Residual spectral baselines
were removed by fitting polynomials in the image domain ({\sc miriad}
task {\sc contsub}).

The multibeam data were calibrated relative to a flux density for
PKS~B1934-638 of 14.9 Jy at the observing frequency. The brightness
temperature conversion factor of 0.80 K Jy$^{-1}$ was established by
an observation of S9 ($T_B=85$ K, Williams 1973). On the same scale,
we measured a brightness temperature for pointing 416
(Stanimirovi\'{c} et al. 1999) in the SMC (RA $00^{\rm h}47^{\rm
m}52.6^{\rm s}$, Dec. $-73\arcdeg 02\arcmin 19\farcs8$, J2000) of
$T_B=133$ K, compared with the 137 K measured by Stanimirovi\'{c} et
al. The 3\% difference is probably due to the different
characteristics of the feeds used in the two observations, and
residual uncertainties in absolute bandpass calibration. The rms noise
in the line-free region of the cube is 27 mK, which is close to the
theoretical value. This corresponds to a column density sensitivity of
$8\times 10^{16}$ cm$^{-2}$ across 1.6 \kms, and $4\times 10^{17}$ cm$^{-2}$
for linewidths of 40 \kms, typical of those in the LMC.

\section{RESULTS}
\label{s:results}

\subsection{Channel Maps}
\label{s:channel}

\begin{figure*}
 \centerline{\psfig{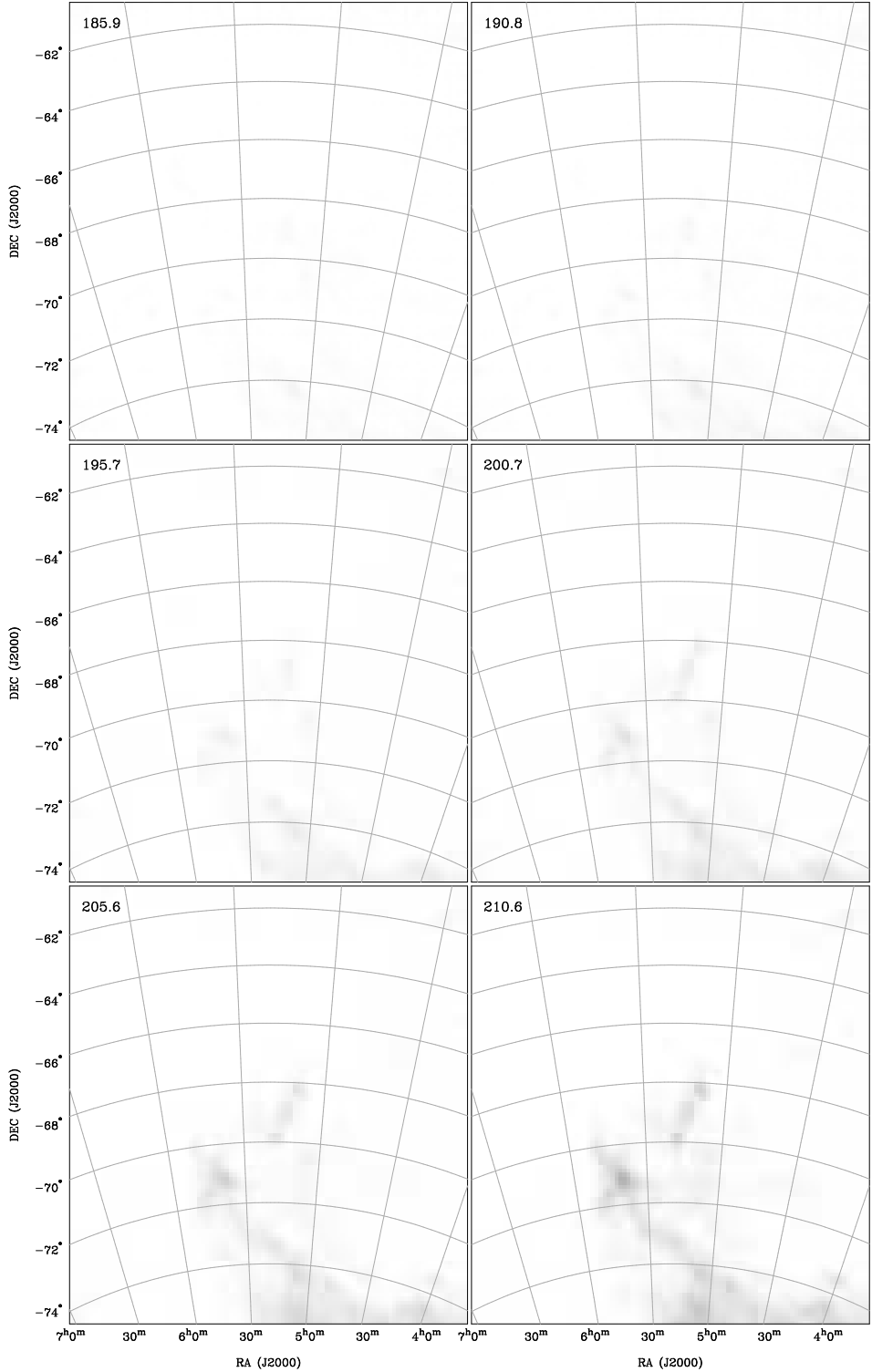}}
 \caption{Channel maps of HI in the LMC formed by averaging six adjacent
   planes of the cube (for display purposes). The resultant velocity
   spacing is 4.92 \kms.  The full intensity range 0 to 83.1 K is
   shown, with a square root transfer function. The heliocentric
   velocity of each velocity plane is shown at the top left in \kms.}
 \label{f:chan_maps}
\end{figure*}

\begin{figure*}
 \centerline{\psfig{figure=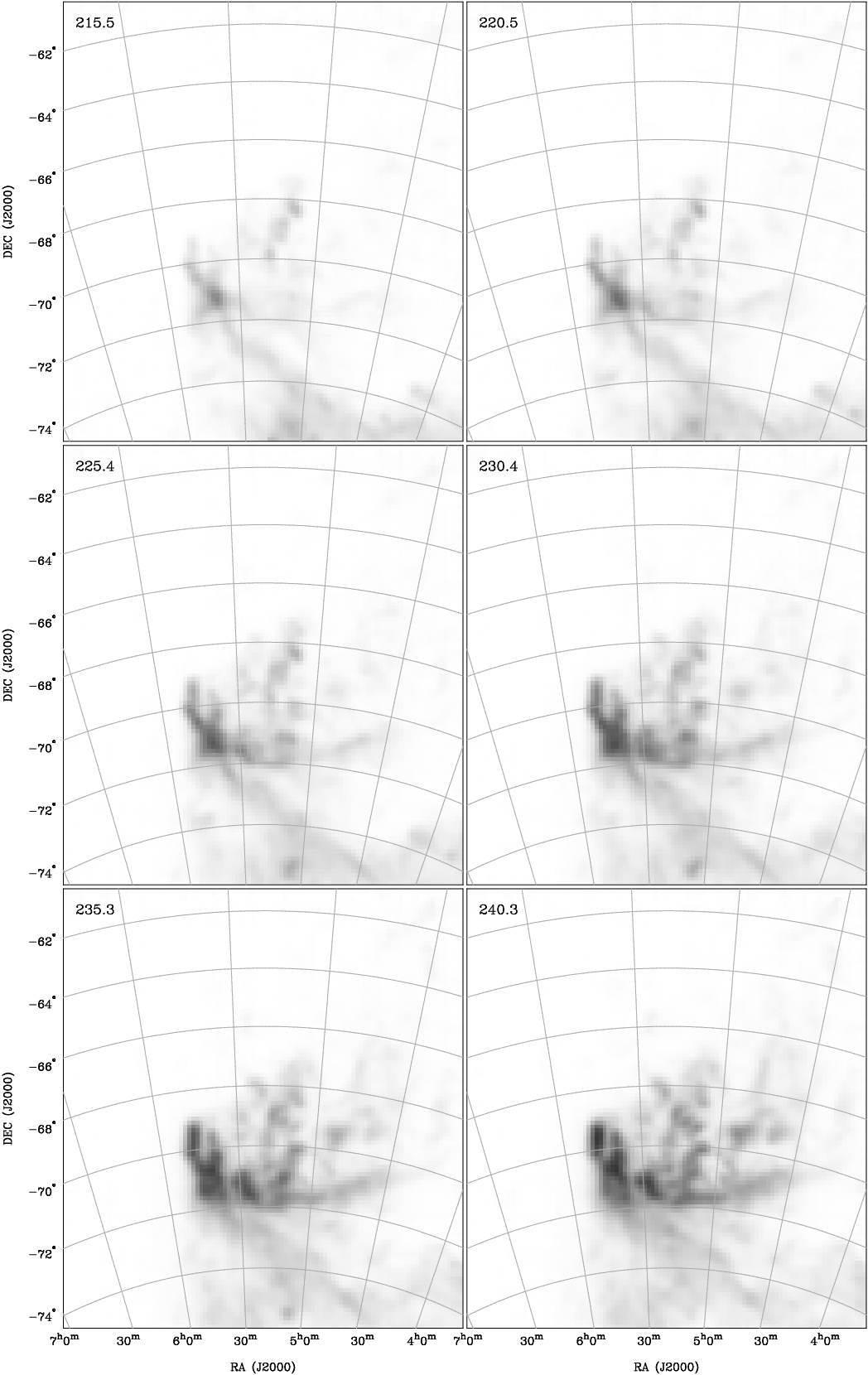,height=20cm}}
 \contcaption{}
\end{figure*}

\begin{figure*}
 \centerline{\psfig{figure=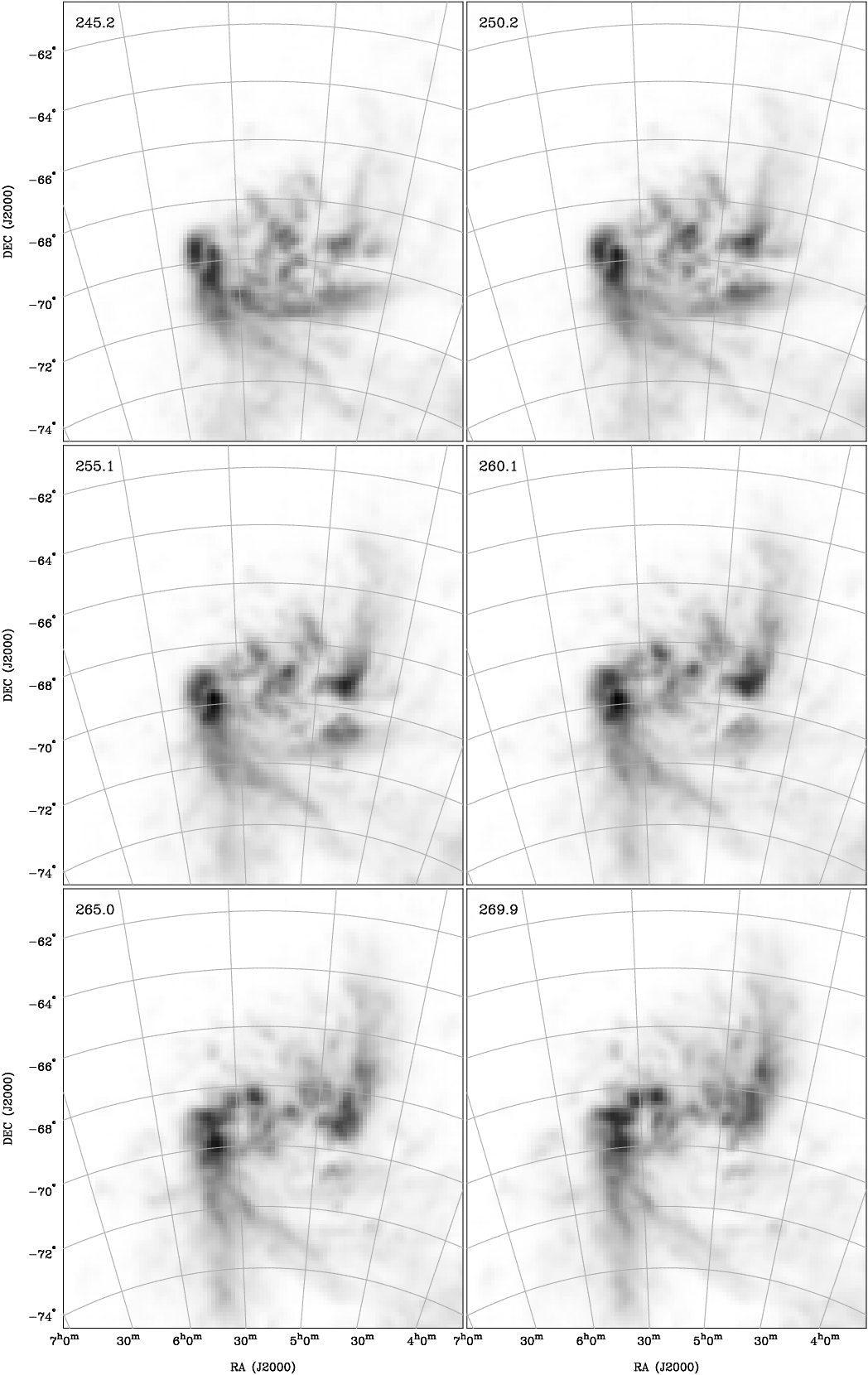,height=20cm}}
 \contcaption{}
\end{figure*}

\begin{figure*}
 \centerline{\psfig{figure=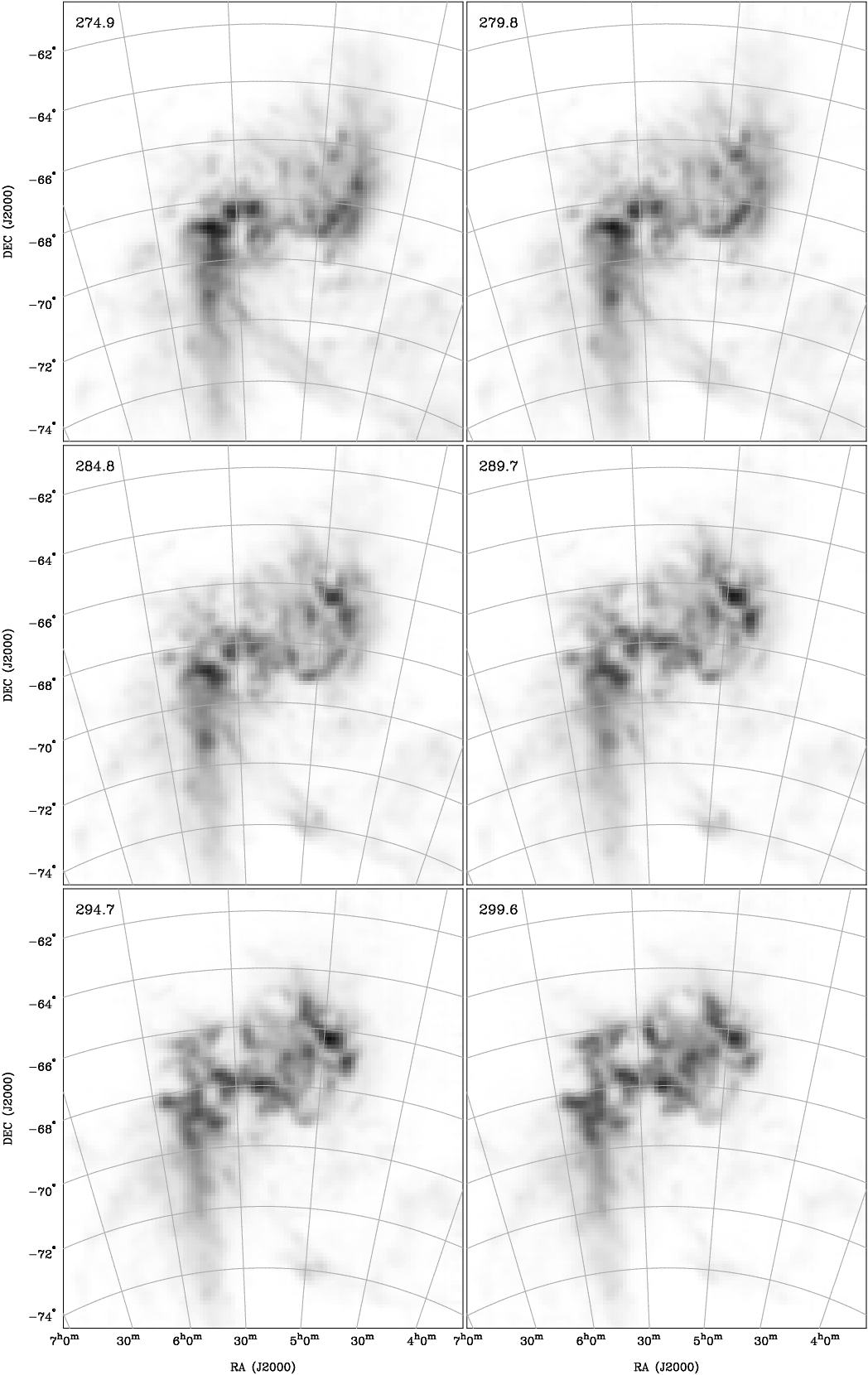,height=20cm}}
 \contcaption{}
\end{figure*}

\begin{figure*}
 \centerline{\psfig{figure=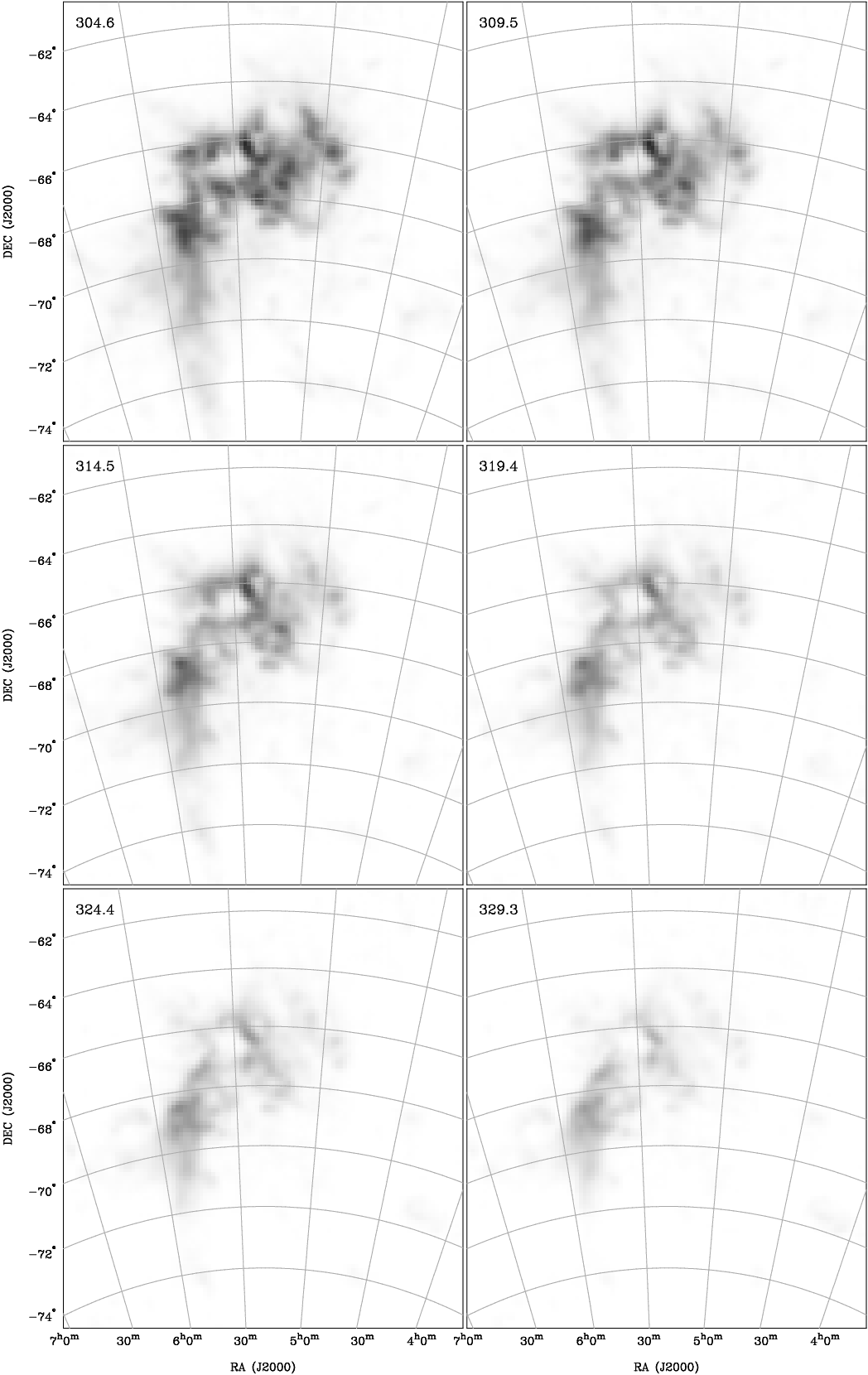,height=20cm}}
 \contcaption{}
\end{figure*}

\begin{figure*}
 \centerline{\psfig{figure=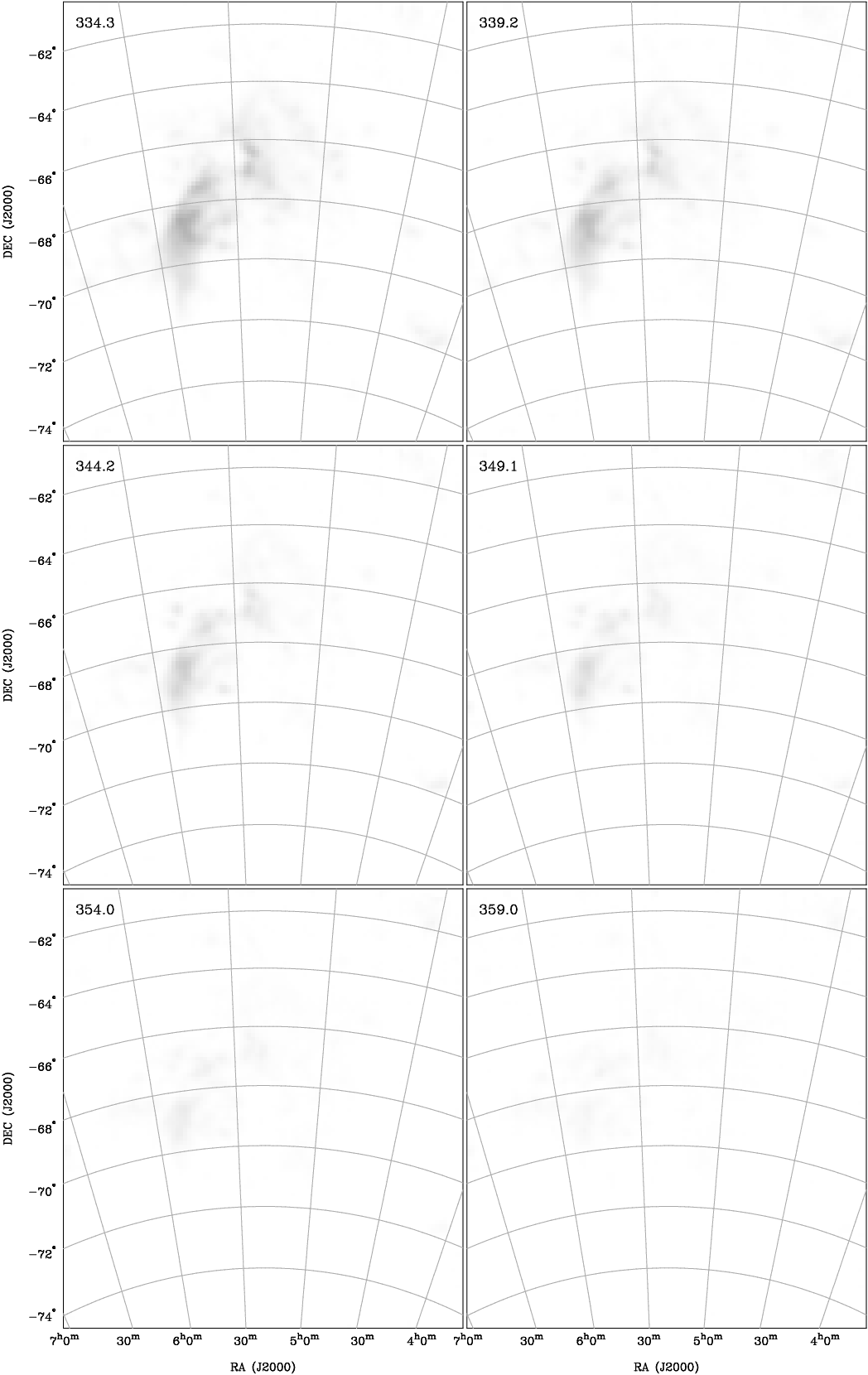,height=20cm}}
 \contcaption{}
\end{figure*}

The area covered by the present observations is shown in Fig.~\ref{f:arms}
together with some prominent features, referred to later in the text.
Relevant directions and proper motion vectors to the Galaxy and other
parts of the Magellanic system are also plotted.

Channel maps, formed by averaging groups of six channels are shown in
Fig.~\ref{f:chan_maps}. Maps between heliocentric velocities of
185.9 and 359.0 \kms\ are shown, spaced by $6\times 0.82 = 4.92$ \kms.
Although this velocity range covers the main body of \HI\ emission in
the LMC, the emission does extend in a continuous manner down to $\sim
100$ \kms\ and up to $\sim 425$ \kms, but at faint levels (see 
Section~\ref{s:hvgas}). There appears to be a small, but clean separation
between \HI\ in the LMC and \HI\ in the Galaxy which extends from
$\sim 90$ \kms\ through to $\sim -50$ \kms. The Galactic component, important
for extinction and photoelectric absorption estimates, is discussed in
Section~\ref{s:galaxy}

The main feature in the channel maps between 186 and 211 \kms\ is the
arm (hereafter arm `B') of the LMC noted before in the Parkes
observations by McGee \& Newton (1986), the ATCA observations of Kim
et al. (1998a), and in the HIPASS map presented by Gardiner, Turfus
\& Putman (1998). Arm B appears to be a tidal feature which directly
connects the LMC with the Magellanic Bridge joining the LMC and SMC.
We discuss this further in Section~\ref{s:tidal}.

Between 220 and 240 \kms, the southern part of the main body of the
LMC becomes visible. The main body appears bounded by another arm
(hereafter arm `S') curving from RA $05^{\rm h} 30^{\rm m}$, Dec.
$-71\degr 30\arcmin$ to RA $04^{\rm h} 20^{\rm m}$, Dec. $-70\degr
00\arcmin$ (J2000), and by a `figure-of-eight' structure associated
with the supergiant shell LMC2 (Points et al. 2000) and the gas complexes to
the south of 30 Doradus (Mochizuki et al.  1994; Blondiau et al. 1997)
which run from RA $05^{\rm h} 41^{\rm m}$, Dec.  $-72\degr 00\arcmin$
to RA $05^{\rm h} 45^{\rm m}$, Dec. $-69\degr 00\arcmin$ (J2000).

Between 260 and 280 \kms, the LMC begins to look remarkably like a
barred spiral galaxy (e.g. NGC~1365). Two arms open out at
approximately Dec. $-69\degr$, the one at RA $04^{\rm h} 45^{\rm m}$
(hereafter arm `W') extending north for 5\degr, and the one at RA
$05^{\rm h} 40^{\rm m}$ (hereafter arm `E') extending south for 6\degr
to at least the limit of the map.  The two arms are connected by \HI\ 
but not in a structure which looks like a bar, nor a structure which
is similar in position or position angle to the optical bar.  Arm E
points directly towards the Leading Arm cloud at RA $05^{\rm h}
28^{\rm m}$, $-80\degr 15\arcmin$ (J2000), $v_{LSR}=323$ \kms\
($\ell=292\fdg4$, $-30\fdg2$, $v_h=333$ \kms) shown in Putman et al.
(1998). This cloud lies 11\degr\ (10 kpc, in projection) south of the
start of arm E, and 4\degr\ beyond the edge of the present map. Deep
HIPASS data show a continuous connection between the Leading Arm and
Arm E at velocities between 260 and 300 \kms, heliocentric.
Throughout the velocity range from 260 and 280 \kms, arm B also
remains visible, showing that this gas is very disturbed and has
velocity extent of at least 100 \kms.

Above 300 \kms, the northern half of the main body of the LMC is
visible, and is dominated by: (a) the supergiant shell LMC4 (Meaburn
1980, Domg\"{o}rgen, Bomans \& de Boer 1995); (b) the supergiant shell
LMC SGS 6 (Kim et al. 1999); and (c) the linear arm E, discussed
above.

\newpage

\subsection{Integrated \HI\ Maps and the Morphology of the LMC}
\label{s:morphology}

\begin{figure*}
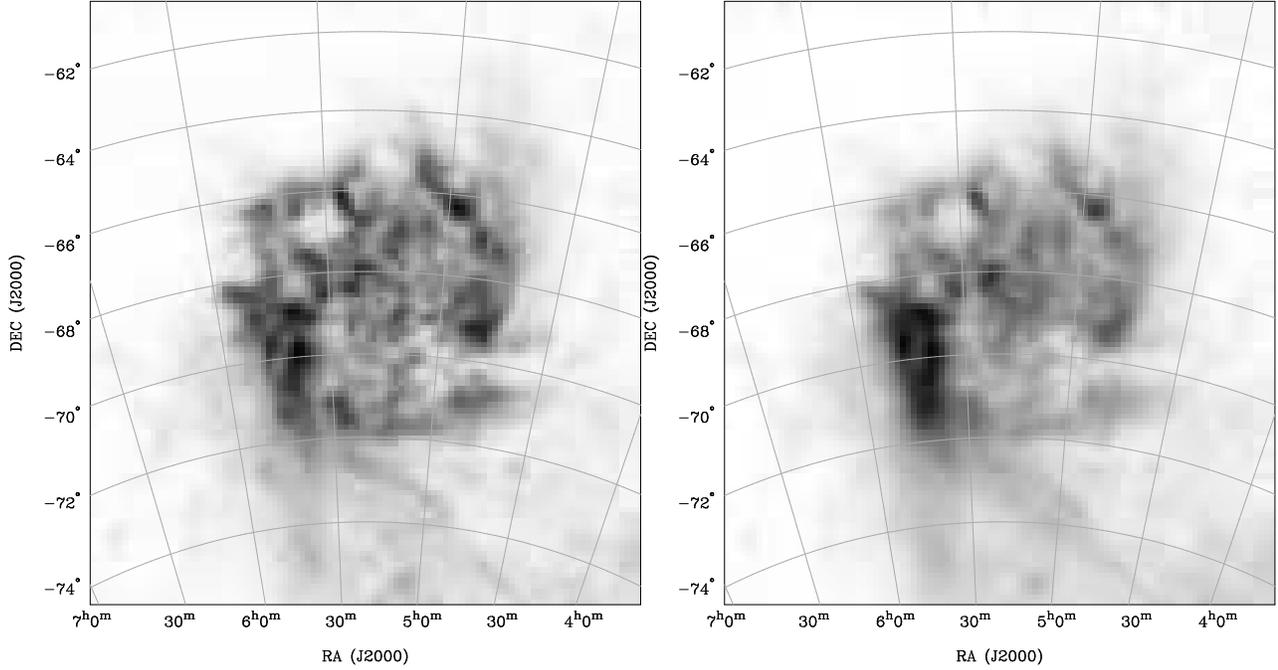

 \centerline{\psfig{figure=fig3a.eps,width=\columnwidth}\psfig{figure=fig3b.eps,width=\columnwidth}}
 \caption{(Left) Peak brightness-temperature image of the LMC showing, for
   each position, the maximum value of $T_B$ in the heliocentric
   velocity range 100 to 430 \kms. The full intensity range 0 to
   83.1~K is shown. (Right) Column density image over the same
   velocity range, formed by summing all emission brighter than 0.08~K
   (3-$\sigma$). The full column density range 0 to $5.6\times10^{21}$
   cm$^{-2}$ is shown. Both images use a square root transfer
   function.}

 \label{f:Tpc}
\end{figure*}

\begin{figure}
 \centerline{\psfig{figure=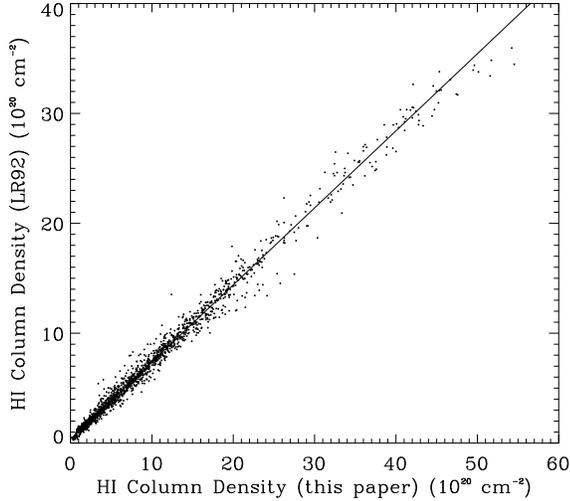,width=\columnwidth}}
 \caption{A pixel--pixel comparison of \HI\ column densities from 
   Luks \& Rohlfs (1992) with the present data. The slope of the fit
   (least absolute deviation) is 0.70. For all points, the mean
   absolute deviation along the $y$-axis is $4.3\times10^{19}$
   cm$^{-2}$. For column densities $<10^{21}$ cm$^{-2}$, the mean
   absolute deviation is $2.9\times10^{19}$ cm$^{-2}$. Pixels at the
   highest column densities fall below the regression line. This may
   reflect the slightly different angular resolutions of the two
   surveys.}

 \label{f:lr92comp}
\end{figure}

\begin{figure}
 \centerline{\psfig{figure=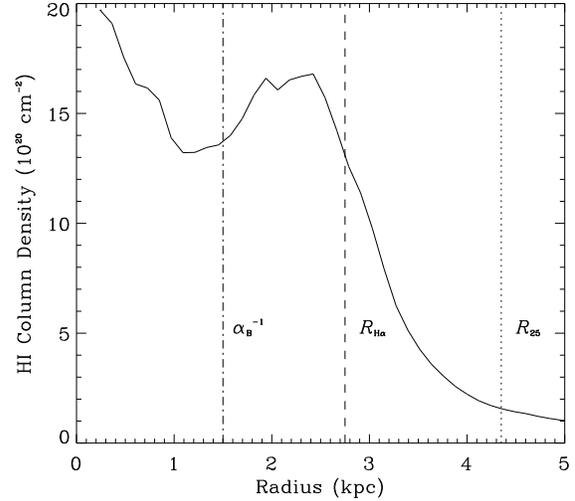,width=\columnwidth}}
 \caption{Azimuthally averaged \HI\ column density profile of the LMC in units
   of $10^{20}$ cm$^{-2}$.  The dynamical centre at RA $05^{\rm h}
   17\farcm6$, Dec.  $-69\degr 02\arcmin$ (J2000) (Kim et al. 1998a)
   has been used, and a distance to the LMC of 50 kpc has been
   assumed. The mean column density is highest near the centre of the
   LMC, and the disk of the LMC appears to be limb-brightened. Various
   optical disk radii are marked: the $B$ scale length,
   $\alpha_B^{-1}$; the H$\alpha$ radius; and the radius at
   $\mu_B = 25$ mag arcsec$^{-2}$, $R_{25}$.  See Table~\ref{t:masses} for
   details.}

 \label{f:hiprofile}
\end{figure}

\begin{figure}
 \centerline{\psfig{figure=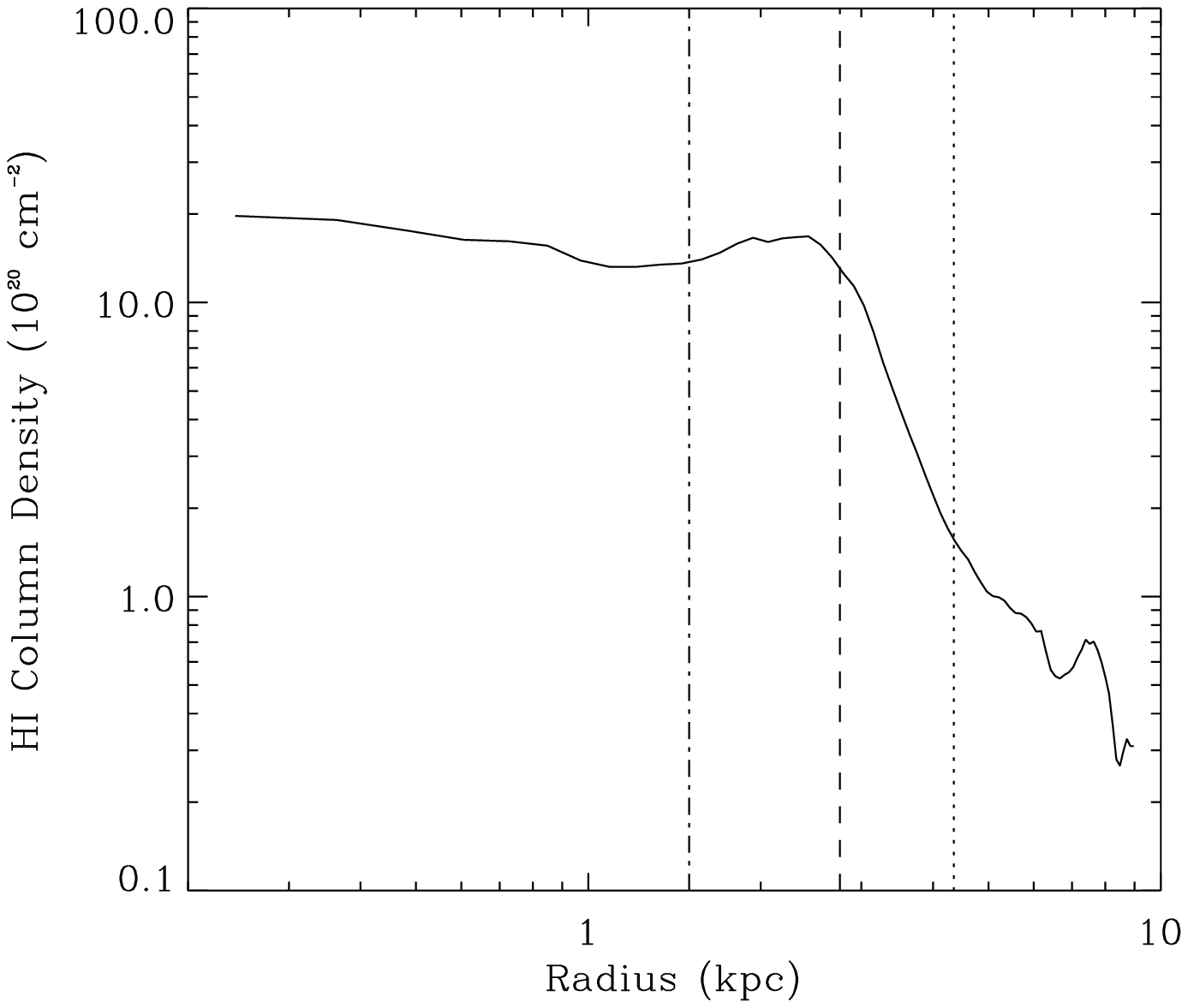,width=\columnwidth}}
 \caption{Azimuthally averaged \HI\ column density profile of the LMC in units
   of $10^{20}$ cm$^{-2}$. As Fig.~\ref{f:hiprofile}, but plotted
   logarithmically and with an extended range of radii. The vertical lines
mark the optical diameters as detailed in Fig.~\ref{f:hiprofile}.}

 \label{f:loghiprofile}
\end{figure}

The peak brightness-temperature image and the column density image of
the LMC are shown in Fig.~\ref{f:Tpc}. The peak brightness
temperature is 83.1 K at RA $05^{\rm h} 39^{\rm m} 22^{\rm s}$, Dec.
$-69\degr 51\arcmin 13\arcsec$ (J2000), a position slightly south of
N159 and the 30 Doradus \HII\ complex. The peak column density of
$5.6\times10^{21}$ cm$^{-2}$ lies at the same position. Both values
are $\sim 50\%$ higher than the values in the Luks \& Rohlfs (1992,
hereafter LR92) data (53 K and $3.6\times10^{21}$ cm$^{-2}$,
respectively). Part of this difference is resolution. The present data
were obtained with the multibeam receiver which has a mean beamwidth
of 14\farcm1, broadened to 16\farcm4\ after scanning and gridding
effects are taken into account (Section~\ref{s:obs}). The LR92 data
were taken with a feed with a beamwidth of $\sim 15\arcmin$ on an
undersampled grid of spacing 12\arcmin\ and interpolated onto a grid
with similar spacing, presumably resulting in an effective resolution
of $\ga 20\arcmin$. However, there must also be a calibration
difference as the \HI\ mass measured here is $\sim 30$ per cent higher
than that of LR92 (Section~\ref{s:himass}). A pixel-by-pixel
comparison of the LR92 column densities and the column densities in
this paper shows that this is the case (Fig.~\ref{f:lr92comp}). The
column density ratio is $N_{\rm HI}/N_{\rm HI}({\rm LR92})=1.43$. This
sizeable calibration anomaly has been noted before (Blondiau et al.
1997 rescale LR92 temperatures by 1.5), and seriously affects use of
LR92 column densities. As confirmed in Luks (1991), LR92
base their calibration on an earlier paper (Rohlfs et al. 1984). This
paper quotes column densities and temperatures based on antenna
temperature, $T_A$ rather than brightness temperature, $T_B$ (their
equation~5 and legend to Table~1). They further state that their \HI\ 
column density ``can be converted approximately into true column
density by multiplication with $1/\eta_{mb}=1.25$''. Assuming this factor 
has been neglected in LR92, the residual
calibration difference then appears to be $1.43/1.25=1.14$. This may
be explained by the high main beam efficiency (or low antenna
efficiency) measured by Rohlfs et al.  They measure a ratio
$T_B/S_{\nu}=0.775$ K Jy$^{-1}$ for the Parkes telescope. In contrast, other
measurements with the same single-beam, hybrid-mode feed suggest
$T_B/S_{\nu}=0.85$ to 0.93 K Jy$^{-1}$ (Davies, Staveley-Smith \& 
Murray 1989, 
Stanimirovi\'{c} et al. 1999), a similar factor ($1.12\pm0.05$) higher.

Our peak column density of $5.6\times10^{21}$ cm$^{-2}$ is also higher
than that of McGee \& Milton (1966) who quote $4.0\times10^{21}$
cm$^{-2}$.  This is probably a resolution difference because, as noted
in Section~\ref{s:himass}, their total \HI\ mass is virtually
identical to ours. The ATCA data of Kim et al. (1998a) at 1\arcmin\ 
resolution show a higher peak brightness temperature of 106 K,
increasing to a true value of 138 K when combined with the present
multibeam data (Kim et al. 2002).

The main features of the \HI\ distribution in  Fig.~\ref{f:Tpc} are:

\begin{itemize}

\item A well-defined, nearly circular disk forms the main body of the
  LMC, indicating a nearly face-on inclination if it is assumed that
  the LMC has an intrinsically circular disk. This agrees with other
  recent values based on this assumption (e.g. 22\degr-26\degr\ 
  Weinberg \& Nikolaev 2001; $22\pm6\degr$, Kim et al. 1998a), though
  not with values based in direct distance determination
  ($42\degr\pm7\degr$, Weinberg \& Nikolaev 2001; $35\degr\pm6\degr$, van der
  Marel \& Cioni 2001). The influence of tidal forces and non-circular
  motions in the outer parts of the disk doubtless contribute to this
  disagreement. van der Marel (2001) deduces an {\it intrinsic}
  ellipticity of 0.31 for the outer disk at near-infrared wavelengths.
  The bulk of the LMC \HII\ regions (Kennicutt et al.  1995, Kim et
  al. 1999) are contained within the gaseous disk.

\item The body of the LMC is punctuated by large holes, and has a
  general mottled appearance. The main \HI\ gaps are at RA $05^{\rm h}
  32^{\rm m}$, Dec. $-66\degr 45\arcmin$ (J2000), corresponding to LMC
  4 (Meaburn 1980) and, in the column density image, the large
  east-west gap centred at RA $05^{\rm h} 00^{\rm m}$, Dec. $-70\degr
  12\arcmin$ (J2000) and bounded by LMC 2 and arm S. This void
  includes LMC 8, LMC SGS 4 (Meaburn 1980, Kim et al. 1999) but is
  substantially larger. This void is discussed in Section~\ref{s:hvgas} and
  the population of \HI\ holes is further discussed in
  Section~\ref{s:holes}.

\item The body of the LMC exhibits limb-brightening in \HI\ as shown
  in the surface density profile of Fig.~\ref{f:hiprofile}. The
  azimuthally-averaged surface density peaks at $2\times 10^{20}$
  cm$^{-2}$ near the dynamical centre, but also again at radius
  2.2 kpc where it reaches $1.7\times 10^{20}$ cm$^{-2}$. The
  limb-brightening is accentuated by arm E and the body of gas near
  LMC~2 and 30 Doradus in the south-east, especially in the column
  density image, and by arms S and W in the south and west,
  respectively. The column density increase in the south-east is
  sometimes identified as compression arising from the proper motion
  of the LMC through the tenuous halo of the Milky Way (e.g. de Boer
  et al. 1998).

\item Diffuse gas is present around much of the LMC, especially at
  pa's from 90\degr\ to 330\degr. At 5 kpc radius,
  Fig.~\ref{f:hiprofile} shows that the mean column density remains
  $1\times 10^{20}$ cm$^{-2}$. A logarithmic version of the surface
  density profile is shown in Fig.~\ref{f:loghiprofile}. This shows
  that, although there a rapid decrease in column density beyond 2.5
  kpc, there is no cutoff. The surface density decrease is
  approximated by $\Sigma(\HI) \propto r^{-3.3}$. This is shallower than
  the exponential profile which seems to characterise the late-type dwarf
  galaxies studied by Swaters (1999).

\item The main arms B and E both emanate, at different
  velocities, from the south-east of the LMC. These arms appear to be
  associated with much of the diffuse gas in the southern half of
  Fig.~\ref{f:Tpc}. These arms are unlikely to be coplanar (see
  Section~\ref{s:tidal}). Arm S is also clearly visible, and appears
  to be associated with some of the diffuse gas in the west. Arm W is
  less visible and more curved in Fig.~\ref{f:Tpc} than in the
  channel maps and may therefore be a superposition of several
  components. As has been noted by several authors recently (Kim et
  al. 1998a; Gardiner et al. 1998) that, although disturbed, the LMC
  has distinct spiral features. This is in contrast with the
  near-infrared map of the stellar distribution (van der Marel 2001)
  which, aside from the bar and bar-related arm-like features, is
  remarkably uniform.

\end{itemize}

The main features in the outer body of the LMC are summarised in
Fig.~\ref{f:arms}.

\subsection{Spatially Integrated Properties of the LMC}
\label{s:himass}

\begin{table*}
\caption{Global properties for the LMC based on distance of 50 kpc.}
\begin{tabular}{llrl}
Parameter & Units & Value  & Reference \\
\hline

Total \HI\ mass, $M_{\rm HI}$ & M$_{\sun}$& $4.8\times 10^8$ & This paper \\
Isodensity mass $(>10^{20} {\rm cm}^{-2})$ & M$_{\sun}$ &
   $4.6\times 10^8$ & This paper \\
Disk \HI\ mass $(<3.5 {\rm kpc})$ & M$_{\sun}$ & $3.8\times 10^8$ &
    This paper\\
\HI\ diameter $(10^{20}\, {\rm cm}^{-2})$ & kpc & 10.2 &
    This paper\\
\HI\ diameter, $D_{\rm HI}\, (1\, M_{\sun}$ pc$^{-2})$ & kpc & 9.3 &
    This paper\\
 & & & \\
Heliocentric velocity, $V_{50}$   & \kms & 273 & This paper \\
Velocity width, $W_{50}$ & \kms &  80 & This paper \\
Velocity width, $W_{20}$ & \kms & 102 & This paper \\
 & & & \\
Foreground extinction, $<E_{B-V}>$ & mag & 0.06 & This paper \\
 & & & \\

Total dynamical mass ($<4$ kpc), $M_T$ & M$_{\sun}$& $5\times 10^9$ & Kim et al. (1998a), Alves \& Nelson (2000) \\
Molecular mass & M$_{\sun}$&  $(4-7)\times 10^7$ & Fukui et al. (1999) \\
Blue luminosity$^a$, $L_B$ & L$_{\sun}$ & $2.3\times 10^9$ & de Vaucouleurs et al. (1991)\\
Blue diameter, $D_{25}$ & kpc  & 8.7 & Bothun \& Thompson (1988), de Vaucouleurs et al. (1991)\\
Blue scale length, $\alpha_B^{-1}$ & kpc  & 1.5 & Bothun \& Thompson (1988)\\
H$\alpha$ luminosity & erg s$^{-1}$ & $2.7\times10^{40}$ & Kennicutt et al. (1995)\\
H$\alpha$ diameter &  kpc & 5.5 & Kim et al. (1999)\\
Star Formation Rate, SFR & M$_{\sun}$ yr$^{-1}$  & 0.26 & Kennicutt et al. (1995) \\
 & & & \\
$M_{\rm HI}/M_T$ ($<4$ kpc) &   & 0.08 &  \\
$M_{\rm HI}/L_B$  &  M$_{\sun}/$L$_{\sun}$ & 0.21  & \\
$(M_{\rm HI}+M_{\rm He})/$SFR & yr & $2.4\times 10^9$  & \\
 & & & \\
$D_{\rm HI}/D_{25}$ &  & 1.1  & \\
$D_{\rm HI}/\alpha_B^{-1}$ &  & 6.2  & \\

\hline
\end{tabular}

$^a$ Based on $B_T^{\circ}=0.57$ (de Vaucouleurs et al. 1991) and
$M_B(\sun)=5.50$ mag (Lang 1991).

\label{t:masses}
\end{table*}

\begin{figure}
 \centerline{\psfig{figure=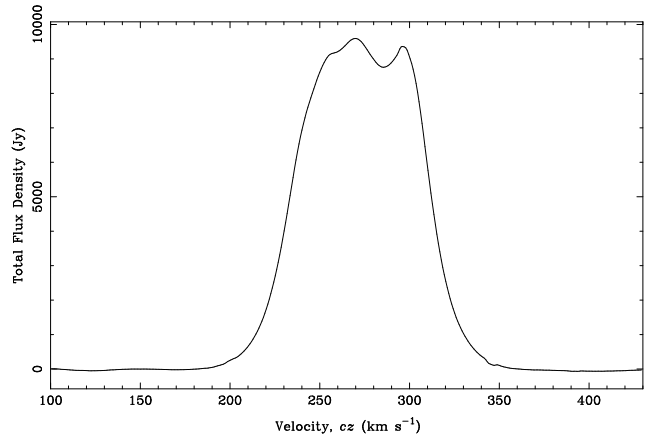,width=\columnwidth}}
 \caption{A global \HI\ profile of the LMC. The spectrum has been spatially
   integrated in a rectangular region of dimensions 9\fdg4 $\times$
   12\fdg7 (RA and Dec., respectively) centred on RA $05^{\rm h}
   17^{\rm m} 36^{\rm s}$, Dec. $-69\degr 02\arcmin$ (J2000). Global
   parameters are summarised in Table~\ref{t:masses}.}
 \label{f:spectrum}
\end{figure}

Assuming a distance of 50 kpc and optically thin emission, the column
density image in Fig.~\ref{f:Tpc} can be integrated to give a total
\HI\ mass for the LMC, $M_{HI} = (4.8\pm 0.2)\times 10^8$ M$_{\sun}$
(Table~\ref{t:masses}). McGee \& Milton (1966) quote a similar value,
$5.0\times 10^8$ M$_{\sun}$ (adjusted to the distance scale used
here), though they applied an optical depth correction based on a spin
temperature of 200~K. As already noted, LR92 quote a smaller value,
$3.1\times 10^8$ M$_{\sun}$. Their survey was limited in spatial
extent.  When our image is integrated over the same area, the mass
remains 30 per cent higher than quoted by LR92. Possible reasons for
this have already been given. It is worth mentioning that our new mass
estimate makes the LMC \HI\ mass higher than that of the SMC which is
$4.2\times 10^8$ M$_{\sun}$ (Stanimirovi\'{c} et al. 1999).  The
isodensity ($>10^{20}$ cm$^{-2}$) mass and disk mass (radius $<3.5$ kpc)
of the LMC are also listed in Table~\ref{t:masses}.

For comparison with other late-type galaxies, the spatially-integrated
\HI\ spectrum of the LMC is shown in Fig.~\ref{f:spectrum}. The area
integrated in this figure is centred on the dynamical centre
of Kim et al. (1998a). The heliocentric velocity (mean and $V_{50}$) is
273 \kms, similar to the Kim et al. (1998a) kinematic value of 279
\kms\ and the LR92 kinematic value of 274 \kms.  We summarise
velocity and velocity width parameters in Table~\ref{t:masses}.

Table~\ref{t:masses} shows that the ratio of the \HI\ to total mass in
the LMC disk is 8\%, the \HI\ mass to blue luminosity ratio is 0.21
M$_{\sun}/$L$_{\sun}$, and the star formation timescale ($M_{\rm
  gas}/$SFR) is 2.4 Gyr, where we have assumed a 30\% He contribution.
The \HI\ mass is a factor of 10 higher than the condensed molecular mass
estimated by Fukui et al. (1999), and a factor of 60 higher than the diffuse
molecular mass measured by Tumlinson et al. (2002).  
The \HI\ diameter of the LMC at 1
M$_{\sun}$ pc$^{-2}$ is 9.3 kpc and the ratio of the \HI\ diameter to
the optical diameter is $D_{\rm HI}/D_{25}=1.1$, in line with the
mean value for spirals of $1.7\pm0.5$ (Broeils \& Rhee 1997), though
less than the mean value for late-type dwarf galaxies of $3.3\pm1.5$
(Swaters 1999). As its absolute magnitude of $M_B=-17.9$ mag and its
morphology indicate, the LMC has properties somewhat closer to those
of spiral galaxies than to those of late-type dwarf galaxies.

\subsection{Tidal and other Interaction Features}
\label{s:tidal}

In Section~\ref{s:channel}, we referred to arms B, E and W which are
marked in Fig.~\ref{f:arms}. Arm B leads directly into the Magellanic
Bridge where it appears to merge with SMC gas, possibly explaining the
multiple-peaked emission profiles in this region (McGee
\& Newton 1986). The Bridge appears to be tidal in origin and, in the
model of Gardiner \& Noguchi (1996), was formed 0.2 Gyr ago. The existence
of arm B demonstrates the presence of some LMC gas in the Bridge, although
the major component is undoubtedly stripped from the lower-mass SMC.
Arm B consists of at least two separate filaments with a separation of
up to $\sim 0.5\degr$ (see Figs~\ref{f:arms} \& \ref{f:Tpc}). The velocity
differential between the filaments was sufficient to warrant listing
the position RA $04^{\rm h} 58^{\rm m} 36^{\rm s}$, 
Dec. $-73\degr 33\arcmin 57\arcsec$ (J2000) as the centre of the
candidate supergiant shell LMC SGS~1 by Kim et al. (1999).

Arms E and W lead directly south and north, respectively, of their 
starting point in the LMC. As already noted, arm E points to the
beginning of the Leading Arm clouds mapped by Putman et al. (1998).
Although these clouds lie 4\degr\ beyond the edge of the present map,
deep reprocessed HIPASS data (Putman et al. 2002) show a continuous
connection between this point and the Leading Arm at heliocentric
velocities between 260 and 300 \kms. Arm W leads directly north,
extending into the diffuse gas to the north-west at $\sim 270$ \kms.
As noted by Putman et al. (2002) and seen in their Fig.5, 
this gas then seems to bypass the Bridge and makes what appears to be 
a direct connection with the Magellanic Stream..

What causes the arm-like features arms E and W? As with spectacular
systems such as NGC 4038/4039 (the ``Antennae''; Hibbard et al. 2001),
a plausible explanation is again tidal interaction. But in this case,
the tidal force arises from the Galaxy and not the SMC. As
Fig.~\ref{f:arms} shows, the great circle to the Galactic Centre lies
directly south\footnote{ For an LMC/Galaxy mass ratio of 0.01, the
LMC-Galaxy Lagrangian L1 point only lies $\sim 5$ kpc in front of the LMC
(and because of parallax, $\sim 1\degr$ south).}  of the LMC,
therefore the tidal force projects along this line, at more or less
the same position angle as the two arms. The two arms E and W may be
wound up due to the LMC's clockwise rotation (Kroupa \&
Bastian 1997). 

It is unusual that arms B and E appear to emanate from similar points
at the south-east of the LMC, with the former flowing into the Bridge
and the latter flowing into the Leading Arm. The chronology of events
is likely to be that the gaseous tide in the LMC was disturbed by the
SMC's close passage 0.2 Gyr ago, funnelling a portion of the gas near
the LMC-Galaxy Lagrangian L1 point into the Bridge. At the present
time, the tidal force ($\propto M/R^3$) from the Galaxy is likely to
be many times stronger than that from the SMC. For an LMC total mass
of $5\times 10^9$ M$_{\sun}$ (Kim et al. 1998, Alves \& Nelson 2000) a
distance of 50 kpc, an SMC total mass of $1.5\times 10^9$ M$_{\sun}$
and an SMC/LMC separation of 22 kpc (Staveley-Smith et al. 1998), the
tidal force ratio\footnote{As an aside, the predicted tidal force of 
the Galaxy on
the LMC in Milgrom's (1983) theory of modified Newtonian dynamics
(MOND) is more or less the same as the Newtonian prediction. However,
because there is little requirement for dark matter in the SMC even
in the Newtonian model, the tidal force in the MOND regime ($\propto
M/R^2$) will be relatively stronger. Thus the Magellanic system, and
similar multiple systems may be useful laboratories for studying
inertia and gravity at low accelerations).}
is $\sim 30$. However, for an encounter at 7 kpc the force ratio would
be unity.

Current numerical models predict that most of the gas in the
Leading Arm (and the Stream) comes from the SMC, 
(Gardiner \& Noguchi 1996; Li 1999), with the LMC merely serving to
disrupt the Leading Arm which passes in front of it. However, it
seems to be the case that, as with the Bridge, significant LMC gas
is `leaking' into the Leading Arm. Metallicity measurements of Leading Arm
clouds such as HVC 287.5+22.5+240 (Lu et al. 1998) may give clues as to
the ratio of LMC gas to the slightly less-enriched SMC gas.

Cepheid distances (Welch et al. 1987) suggest that the closest part of
the LMC is at pa $77\arcdeg \pm 42\arcdeg$ Recent AGB and RGB stellar
distances give a more accurate near-side pa of $32\arcdeg \pm8\arcdeg$
(van der Marel \& Cioni 2001). Therefore, the north-eastern part of
the disk is the closest to the Galaxy, closest to the L1
point and the most easily perturbed by the SMC. This is not
inconsistent with the point of origin of arms B and E.

A suggested geometry for the LMC tidal features is that arms B and E
both arise from the outer parts of the LMC in the east. The arms
extend southwards where they bifurcate at a position close to the LMC
tidal radius. The low-velocity arm B swings around, probably upwards
out of the plane of the LMC, curves around to the south-west where it
eventually joins the Magellanic Bridge at somewhat larger distances
than the LMC. This feature is evident in the simulations of Li (1999).
The high-velocity arm E extends directly south, probably remaining at
the same Galactocentric distance as the LMC. It then joins the general
Leading Arm gas which mainly arises from the SMC. The large amount of
star-formation in the 30 Doradus region may well be a manifestation of
the tidal shear occurring in the region near the origin of arms B and
E.

However, the question remains as to why the near-infrared stellar
distribution is so smooth in the outer parts of the LMC. van der Marel
(2001) reports no clearly discernable spiral structure at radii out to
9\arcdeg, which is beyond the radius surveyed in this paper. Tidal
distortion must apply to stars as well as gas, and the usual argument
that the \HI\ is well outside the stellar distribution does not apply.
Could it be that the evolved stars in the 2MASS and DENIS images are
not tracing the same thin disk as the \HI? Or has orbit-crossing and
dissipation made the \HI\ density evolve in a non-linear manner? The
significant gas self gravity ($\sim 10$\% of the total mass is in the form
of \HI\ and He; see
Table~\ref{t:masses}) makes the latter a distinct possibility.
Nevertheless, the stellar distribution shows strong
evidence of tides. van der Marel (2001) argues that the
distribution of stars is significantly elongated in the direction of
the Galactic Centre, as predicted by tidal theory.

The diffuse gas in Fig~\ref{f:Tpc}, and labelled in
Fig.~\ref{f:arms} appears to be strongly related to the tidal arms.
The gas in the south-east occurs at similar velocities to arm E. In
the channel maps (Fig.\ref{f:chan_maps}), some of this gas at 290
\kms\ itself falls into diffuse linear features. The more extensive
HIPASS data again shows a connection to the Leading Arm.  This would
tend to argue for a tidal origin for this gas also. Weinberg (2000)
also points out the dramatic effect of the Galaxy on the LMC's stellar
structure through torquing of disk orbits and tidal stripping, and
suggests a mass-loss rate of $3\times10^8$ M$_{\sun}$ per orbit, even
without the SMC. He also predicts tidally stripped stars some tens of
kiloparsecs away from the LMC, which have possibly already been seen
in the 2MASS data (Weinberg \& Nikolaev 2001). With an LMC inclination
of $\sim 30\arcdeg$, and a line-of-nodes at pa $\sim 0\arcdeg$, the
spin vectors of the Galaxy and the LMC are perpendicular, a geometry
not suitable for strong tidal interaction. However, if the higher
values for the inclination and pa suggested by van der Marel (2001)
are correct, then the spin vectors of the Galaxy and the LMC will be
better aligned, and the current interaction will be stronger.

For the diffuse gas in the south-east, it is also worth noting
that ram pressure is a viable alternative.  Any gas
outside the influence of the LMC which is slowed by ram pressure will
tend to drop in its orbit around the Galaxy. By doing this, its
angular velocity increases and it will tend to lead the LMC, unless
dynamical friction on the LMC is more important (Moore \& Davis 1994).
Although Murali (2000) suggests that ram pressure is not a viable
possibility for a remnant as old as the Magellanic Stream (otherwise
there is too much evaporation), this argument does not apply to the
clouds near the LMC.  On the other side of the LMC, the diffuse gas to
the south-west is clearly associated with the Bridge.

As well as having a strong tidal influence on the LMC and SMC, the
Galaxy has left a clear tidal signature on the Sagittarius dwarf
galaxy which is responsible for a trail of stars forming a great
circle on the sky (Ibata et al. 2001). Mart\'{i}nez-Delgado et al. (2001) 
confirm the existence of tidal tails in Ursa Minor,
a dwarf spheroidal galaxy lying at a distance of 70 kpc. Other objects
such as the Carina dwarf spheroidal galaxy also appear to have stellar
distributions which are radially truncated (Majewski et
al. 2000). However, a lack of a clear understanding of the internal
dynamical state of these systems has contributed to an uncertainty about 
whether the truncation can strictly be interpreted as tidal in origin.

The importance of tidal signatures in satellite galaxies lies in their
usefulness in measuring the total mass of the Milky Way, the extent of
its dark halo, and for detecting any sub-structure within the halo.
Present results for the LMC and the Magellanic Stream appear to
confirm the existence of a massive halo around the Galaxy, with the
total mass out to a radius of 50 kpc of $\sim 5\times10^{11}$
M$_{\odot}$. However, the existence of tidal debris around the LMC
complicates interpretation of results from the MACHO experiment
(Alcock et al. 2000) which attempted to measure the fraction of the
Galaxy's dark halo in the form of compact objects. Stars which are
tidally stripped and lie slightly behind the LMC, rather than in a
thin disk, substantially contribute to the frequency of microlensing
events (Weinberg 2000). This implies that the measurement of the
Galactic halo mass in the form of compact objects (currently $\sim
10^{11}$ M$_{\odot}$) should probably be regarded as an upper limit
until the geometry of the halo of the LMC is better understood.

\section{THE HALO OF THE LMC}
\label{s:hvgas}

\begin{figure}
 \centerline{\psfig{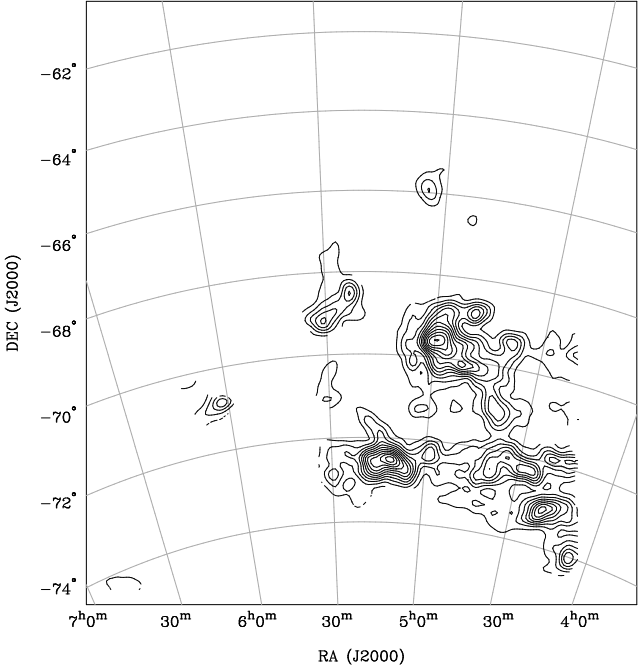}}
 \caption{\HI\ column density contours of high-velocity clouds.
   The heliocentric velocity range 115 to 176 \kms\ is included.
   Contours are in steps of $2\times 10^{18}$ cm$^{-2}$ starting at
   $2\times 10^{18}$ cm$^{-2}$. Noisy regions at the edge of the image
   have been excluded. The map has been smoothed by a Gaussian of FWHM
   12\arcmin, and was derived by applying a smooth mask to the data
   cube in order to isolate those regions with significant column
   density.  Many of the high-velocity clouds shown here are probably
   connected to the LMC.  The peak column density of $2.4\times
   10^{19}$ cm$^{-2}$ occurs at $04^{\rm h} 59.3^{\rm m}$, Dec.
   $-69\degr 35\arcmin$ (J2000) which is the position of an \HI\ void
   and the supergiant shells LMC 8 and LMC SGS 4 (which are also discussed in
   Section~\ref{s:holes}).}
 \label{f:hvccontours}
\end{figure}

\begin{figure}
 \centerline{\psfig{figure=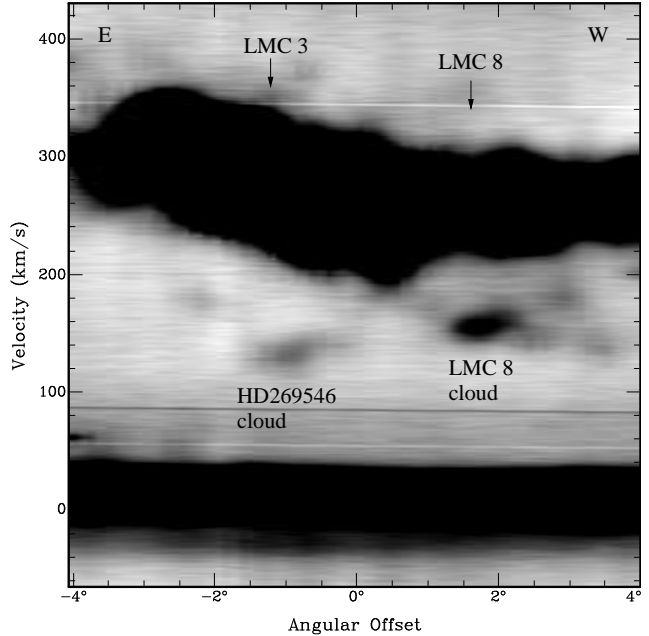,width=\columnwidth}}
 \caption{An \HI\ position-velocity slice centred on $05^{\rm h}
17^{\rm m} 02^{\rm s}$, Dec.  $-68\degr 49\arcmin 19\arcsec$ (J2000) at pa
250\degr. The emission from the LMC (at $\sim 260$ \kms)
and the Galaxy (at $\sim 10$ \kms) has been saturated to reveal the
faint emission from the HD269546 absorbing cloud (Richter et al. 1999)
at $05^{\rm h} 27^{\rm m}$, Dec.  $-68\degr 50\arcmin$ (J2000), 131 \kms\
(heliocentric) and the LMC 8 cloud at $04^{\rm h} 59^{\rm m}$, Dec.  
$-69\degr 35\arcmin$ (J2000), 155 \kms\ (heliocentric). The data has been
smoothed with a Gaussian FWHM of 20\arcmin. The narrow feature at $\sim 85$
\kms\ is an artefact.}
 \label{f:pvslice}
\end{figure}

Because the present study has velocity information, we can attempt to
probe the LMC's halo by searching for \HI\ at anomalous
velocities. Such gas is expected to occur due to outflows from
star-forming regions and stripping of the outer disk due to tidal and
ram pressure forces.  Non-planar occurence of gas and stars has
previously been reported or suggested by Luks \& Rohlfs (1992),
Zaritsky \& Lin (1997), Wakker et al.  (1998) and Graff et al. (2000)
and, as reported by the latter, is an important factor in determining
the self-lensing optical depth of the LMC. Of particular interest are
the HVC complexes below $+180$
\kms, whether they belong to the Galaxy or the LMC. de Boer, Morras \&
Bajaja (1990) use an \HI\ strip scan to conclude that the gas at LSR
velocities $+70$ and $+140$ \kms\ is Galactic in origin. Richter et al.
(1999) concur, suggesting that the $+120$ \kms\ H$_2$ absorption
feature seen against HD269546 is an HVC originating in the disk of
our Galaxy.

Fig.~\ref{f:hvccontours} shows the prominent HVCs (column densities
over $2\times 10^{18}$ cm$^{-2}$) with velocities in the range 115 to
176 \kms\ (heliocentric). They are extensive over much of the
south-west LMC, but extend elsewhere in the field at lower column
densities. Some or most of the emission in the south-west lies projected
between arms B and S, and is likely related to the Bridge and the
tidal interaction with the SMC.  However, the peak column of
$2.4\times 10^{19}$ cm$^{-2}$ projects onto the position of the giant
\HI\ void containing LMC 8 (Meaburn 1980) and LMC SGS 4 (Kim et al.
1999) (see also Sections~\ref{s:morphology} and \ref{s:holes}).  Oey
et al. (2002) show a high-resolution \HI\ mosaic of this region which
also contains the superbubbles DEM L25 and L50. Moreover, the
HD269546 absorbing cloud seen by Richter et al. (1999), which has a
peak column density $1.0\times 10^{19}$ cm$^{-2}$, is at RA $05^{\rm
  h} 27^{\rm m}$, Dec.  $-68\degr 50\arcmin$ (J2000) which corresponds
to the \HI\ void at LMC 3 (Meaburn 1980) and LMC SGS 12 (Kim et al.
1999).  This suggests that this gas has been removed from the LMC disk. A
position-velocity cut across both the HD 269546/LMC 3 and the LMC 8
clouds is shown in Fig.~\ref{f:pvslice}. The latter shows clear
connections with LMC gas.  The former shows a probable connection at
low column density. At other pa's, both complexes also appear to have
gas at velocities higher than the LMC disk. This suggests an
explosive origin for the gas, rather than a tidal or
ram-pressure-stripped origin. However, the kinematics of the
high-velocity gas are not simple, and certainly cannot be modelled as a
simple expanding bubble, or double-sided mushroom cloud.

The association of some high-velocity clouds with the LMC implies that
the outflow velocities, however they are attained, are substantial. In
the case of the HD269546 cloud, the mean velocity differs by 126 \kms\ 
from systemic. In the case of the LMC 8 cloud, the mean velocity
differs by 97 \kms\ from systemic. Outflows at these velocities ought
to be accompanied by X-ray as well as the H$\alpha$ emission already
known to be present. However, the strength depends on the density of
the medium being shocked. Because a substantial \HI\ column is
present in the HD269546 cloud, its temperature is lower than $10^4$ K, but 
higher than the lower limit to the H$_2$ excitation temperature of
$10^3$ K measured by Bluhm et al. (2001). These clouds probably
exist in a cocoon of hotter material. Bluhm et al. (2001) suggest
a neutral hydrogen fraction of 5--20\%, from observations of O and O$^+$.

 The height of the neutral clouds above
the LMC disk can only be speculated. The projected radii of LMC SGS 4
and 12 (Kim et al. 1999) provides a weak lower limit of $\sim 0.5$
kpc. The $\sim30$\degr\ inclination of the LMC possibly provides an upper
limit of a few kpc, otherwise the non-planar gas would project
elsewhere on the LMC disk (though this may indeed be the case for much
of the gas in Fig.~\ref{f:hvccontours}).

\section{\HI\ HOLES}
\label{s:holes}

\begin{figure}
 \centerline{\psfig{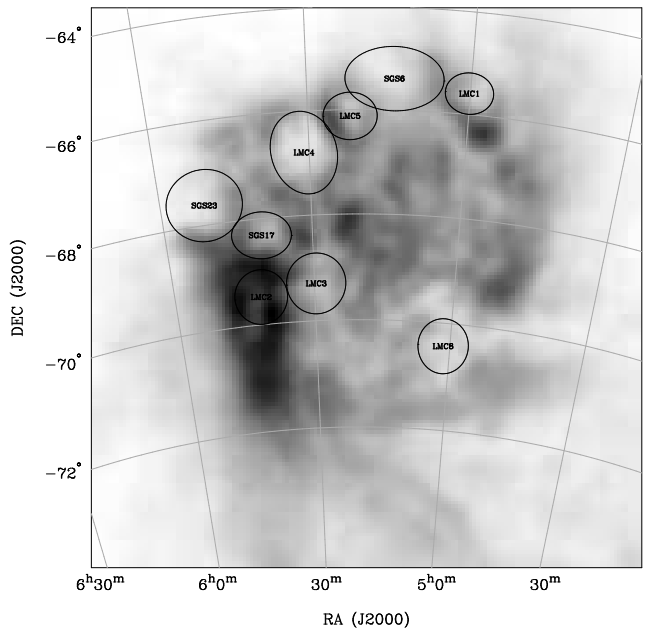}}
 \caption{Supergiant shells from Meaburn (1980) (LMC 1 -- 5 and 8) 
   and Kim et al. (1999) (LMC SGS 6, 17 and 23) are overlayed on the
   \HI\ column density image (slightly enlarged compared with
   Fig.~\ref{f:Tpc}) if they coincide with clear \HI\ holes at the
   resolution of the Parkes data. The shells LMC 1 to 5 and 8 are also
   known as LMC SGS 3, 19, 12, 11, 7 and 4, respectively.}
 \label{f:sgshell}
\end{figure}

\begin{figure*}
 \centerline{\psfig{figure=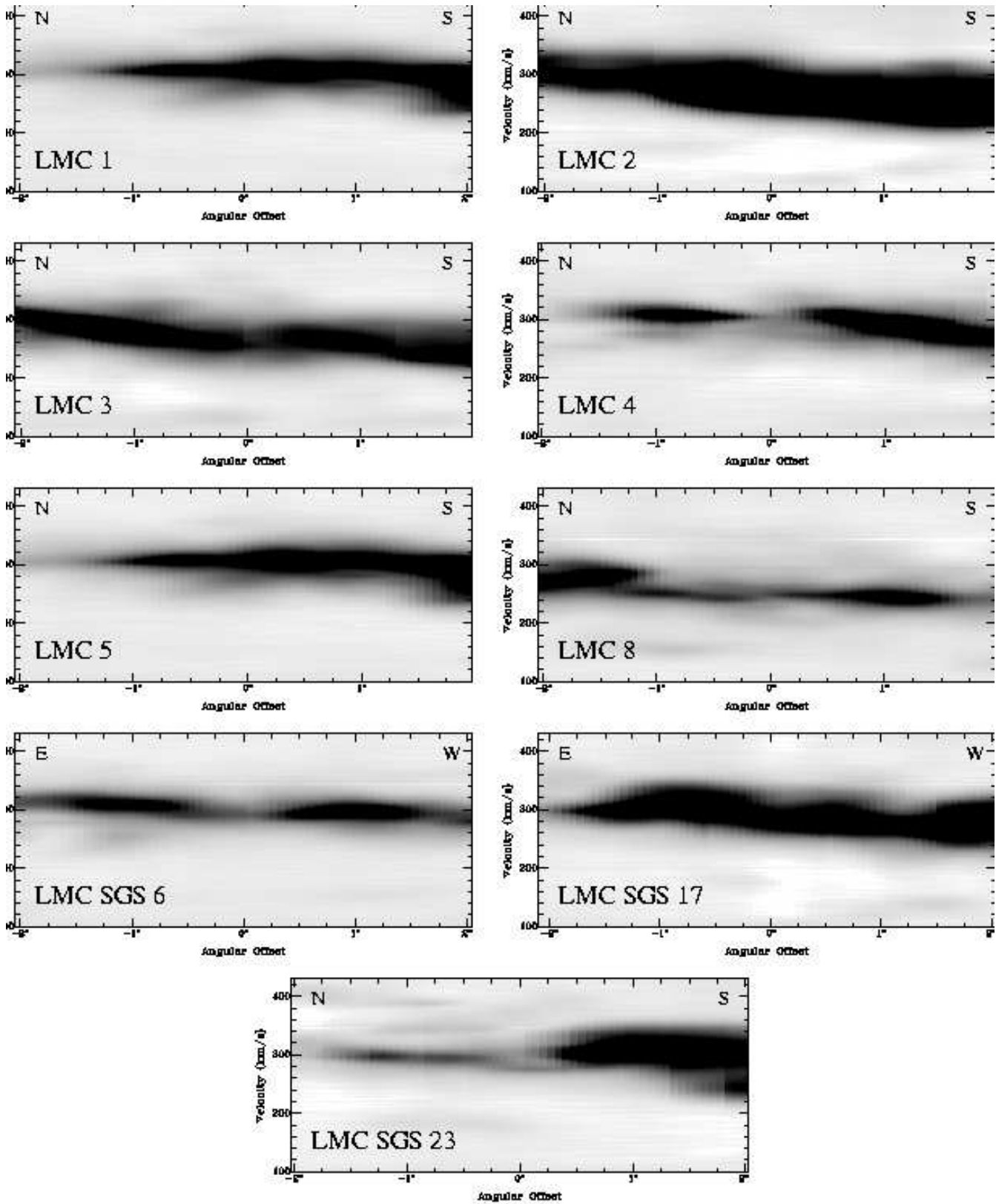,width=\textwidth}}
 \caption{Position-velocity slices across the \HI\ holes marked in 
   Fig.~\ref{f:sgshell}. The slices are generally at pa 0\degr\ 
   except for LMC 4 (10\degr), LMC SGS 6 (90\degr) and LMC SGS 17
   (85\degr). To bring out fainter features, the greyscale is
   saturated at 9 K. High-velocity gas appears to be associated with
   some of the holes (e.g. LMC 4), but in only two cases (LMC SGS 17 
   and 23) is there some evidence for a complete shell. The data
   have been smoothed with a Gaussian FWHM of 20\arcmin.}
 \label{f:PV_mosaic}
\end{figure*}

Several holes, or gaps in the \HI\ column density image in
Fig.~\ref{f:Tpc} are apparent. As originally noted by Westerlund \&
Mathewson (1966), the positions of some holes show a very strong
correlation with stellar associations, \HII\ regions and SNRs.
Six of the nine candidate supergiant H$\alpha$ shells listed by
Meaburn (1980) are clearly associated with holes. These shells are
marked in Fig.~\ref{f:sgshell}. LMC 1, 4 and 5 are well-aligned with an
\HI\ hole; LMC 2 and 3 are more complex than simple circular holes;
and LMC 8 is associated with an \HI\ hole that is at least twice as
large as the H$\alpha$ shell. In addition to the six Meaburn shells
shown in Fig.~\ref{f:sgshell}, Kim et al. (1999) list a further 17
supergiant shells. Some of the largest of these (LMC SGS 6, 17 and 23) are
also clearly visible as \HI\ holes in Fig.~\ref{f:sgshell}.

As discussed by Kim et al. (1999), there is a good association between
\HI\ and \HII\ regions, and some evidence for regions of
star-formation providing direct mechanical input into the expansion of
the shells and therefore the evacuation of the \HI. However, Wada et
al. (2000) point out that thermal and gravitational instabilities can
also lead to the formation of cavities and filaments. In cases like
LMC 4, numerous studies (Dopita et al. 1985, Domg\"orgen et al. 1995,
Efremov \& Elmegreen 1998, Olsen et al. 2001) paint a picture of
progressive star-formation propagating outwards from the centre of the
shell. Ionization, dynamical pressure and conversion to molecules and
stars depletes the \HI, with the remainder forming a rim where the
newest star-formation occurs. In other cases such as LMC 2, the
geometry is more complex and, although energy input by stellar winds
and supernovae (Weaver et al. 1977) may dominate, the simple model of
outwardly propagating star-formation requires modification (Points et
al. 1999, 2000). Finally, in cases such as LMC SGS 6 (Kim et al.
1999), no correlation between \HI\ and anything else is evident. In
these cases, it is possible that an older population of stars whose
initial mass function was skewed towards higher mass stars is
responsible.  However, it is also possible that other dynamical events
are responsible.

For each of the holes marked in Fig.~\ref{f:sgshell}, we have
plotted a position-velocity slice in Fig.~\ref{f:PV_mosaic}. These
slices commonly show a dip in the column density at the position of
the hole (except where the image is saturated), evidence for a
velocity gradient which reflects the rotation of the LMC, and, in some
cases, evidence for high-velocity gas. LMC 2, 4 and LMC SGS 17 and 23
show evidence for multiple velocity components close to the systemic
velocity ($\Delta V \sim 20$ \kms) and which may be construed as
continued slow momentum or pressure-driven expansion of gas away from
the interior of the evacuated shell. In addition, most of the holes
(except LMC 4 and LMC SGS 6) show evidence for higher velocity gas
($\Delta V \sim 100$ \kms) at low column densities (as discussed in
Section~\ref{s:hvgas} for LMC 3 and 8). The origin of this gas and its
definite association with the \HI\ holes remains unclear. However,
expulsion by stellar and supernova shocks, or other explosive events,
remains a good possibility. Further studies of the coronal gas towards such 
regions (e.g. Bomans et al. 1996, Wakker et al. 1998) are desirable.

\section{THE GALACTIC FOREGROUND}
\label{s:galaxy}

\begin{figure*}
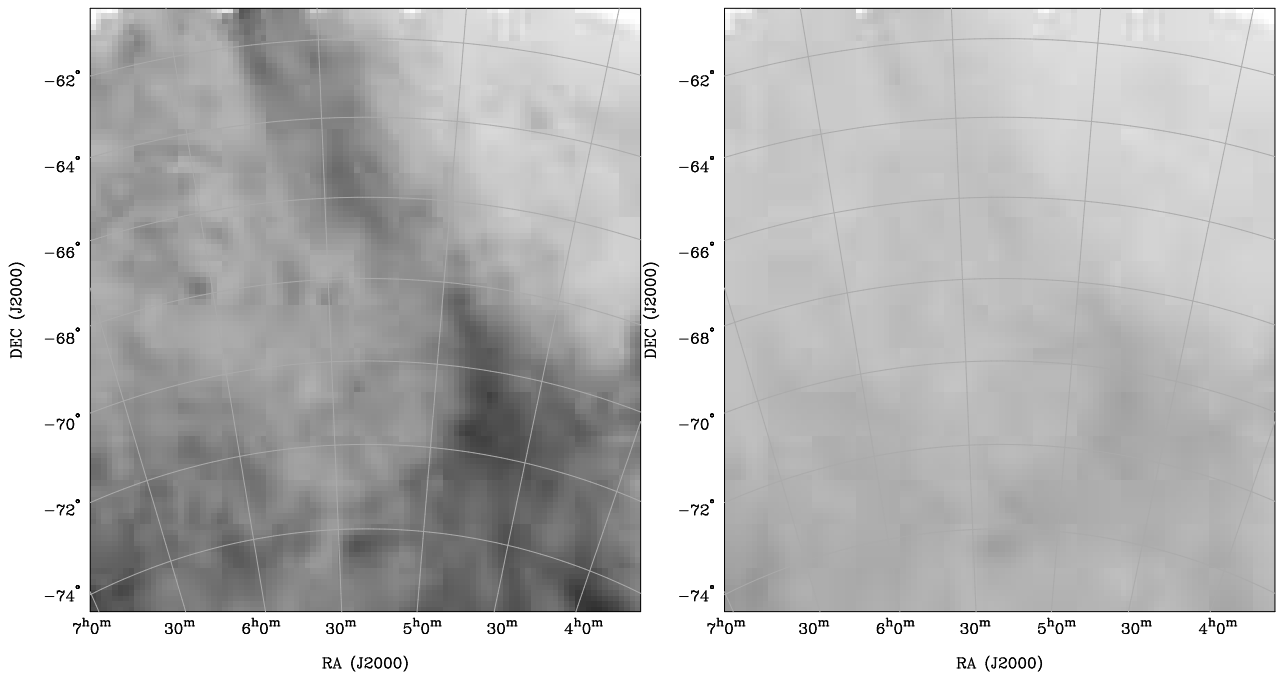

 \centerline{\psfig{figure=fig12a.eps,width=\columnwidth}\psfig{figure=fig12b.eps,width=\columnwidth}}
 \caption{(Left) Peak brightness-temperature image of the Galactic gas in 
   the foreground of the LMC showing, for each position, the maximum
   value of $T_B$ in the heliocentric velocity range $-64$ to 100
   \kms.  The intensity range and transfer function is the same as for
   Fig.~\ref{f:Tpc}, although the peak temperature for the Galactic gas
   is lower, 56.3~K. (Right) Column density image over the same
   velocity range, formed by summing all emission brighter than 0.08~K
   (3-$\sigma$).  The column density range is the same as for
   Fig.~\ref{f:Tpc}, with the maximum column density for the Galactic
   gas being $1.3\times10^{21}$ cm$^{-2}$.}

 \label{f:Gpc}
\end{figure*}

Its closeness, favourable inclination and low internal extinction make
the LMC an ideal object to study in the optical for many purposes.
However, its mean Galactic latitude of $-34\degr$ means that
foreground extinction is of some importance. {\it IRAS} is unable to
separate LMC dust from Galactic dust (although Schwering \& Israel
1991 attempted to incorporate low resolution \HI\ data to isolate the
foreground component). For example, the LMC, SMC and M31 are not
removed from the IRAS/DIRBE extinction maps of Schlegel, Finkbeiner \&
Davis (1998).  Around the outskirts of the LMC, the maps of Schlegel
et al. (1998) show a variation from $E_{B-V}=0.04$ to the north-west
of the LMC to around 0.12 in the south-west, implying that extinction
is significant, and variable. The SMC, lying slightly further from the
Galactic Plane appears to lie in a region with foreground extinction
at the lower end of this range. A strong linear feature, now known to
be associated with Galactic gas and dust also complicates the picture
(de Vaucouleurs 1955, McGee et al. 1986).

\begin{figure*}
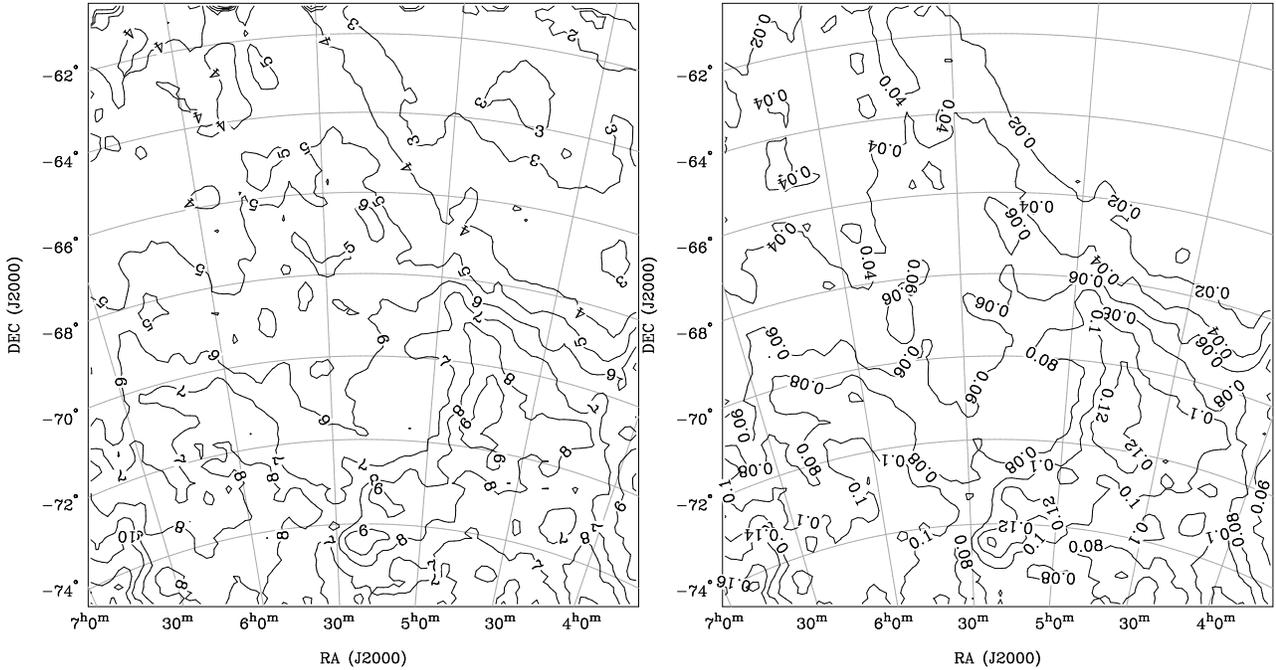

 \centerline{\psfig{figure=fig13a.eps,width=\columnwidth}\psfig{figure=fig13b.eps,width=\columnwidth}}
 \caption{(Left) Contours of \HI\ column density of Galactic foreground
   gas at heliocentric velocities $<100$ \kms. Contours are labelled
   in units of $10^{20}$ cm$^{-2}$; (Right) contours of estimated
   extinction $E_{B-V}$ due to Galactic foreground dust.  Contours are
   labelled in units of mag.}

 \label{f:galcont}
\end{figure*}

Because there appears to be no Galactic gas $>100$ \kms, or LMC gas
$<100$ \kms, there is a clean separation of the Galactic and LMC
components in the \HI\ data.  Fig.~\ref{f:Gpc} shows the peak
brightness-temperature and column density images for the Galactic
component using the same intensity range as the LMC images in 
Fig.~\ref{f:Tpc}.
The Galactic component is smoother with a maximum column density
around a quarter of that in the LMC, and a maximum temperature of
around two-thirds of the LMC value. There is a significant
column-density gradient from north-west to south-east and a
significant filament extending from the north at RA $05^{\rm h}
38^{\rm m}$, Dec. $-62\degr 26\arcmin$ (J2000) to RA $04^{\rm h}
39^{\rm m}$, Dec. $-70\degr 50\arcmin$ (J2000) in the south-east.
Contours of \HI\ column density are shown in Fig.~\ref{f:galcont}
along with $E_{B-V}$ contours based, for reference, on equation (7) of
Burstein \& Heiles (1978). 

Over the disk of the LMC, the mean Galactic extinction is
$<E_{B-V}>=0.06$ mag with a range between 0.01 and 0.14, and an rms of
0.02 mag. The extinction gradient is from north-west to south-east,
with a systematic variation of $\Delta E_{B-V} \approx 0.1$ mag over
the full field. Although significant in $B$ ($\Delta A_B \approx 0.4$
mag), the variation in $H$ is small ($\Delta A_H \approx 0.06$ mag)
and should only affect the details of infrared structural results
(Weinberg \& Nikolaev 2001, van der Marel \& Cioni 2001) rather than
modify their overall conclusions. As pointed out by Schwering \&
Israel (1991), 30 Doradus has a low foreground extinction
($E_{B-V}=0.05$).  The LMC values agree with: the mean of
$0.06\pm0.02$ mag and range of 0.00 to 0.15 mag from the cool star
data of Oestreicher, Gochermann \& Schmidt-Kaler (1995) (see also
Zaritsky 1999); a mean of 0.06 mag from a combination of polarisation
and \HI\ data by Bessell (1991); and the range 0.07 to 0.17 mag
suggested by Schwering \& Israel (1991) from a combination of \HI\ and
{\it IRAS} data.

Finally, it should be emphasised that the \HI\ data alone has
zero-point problems due to a combination of spillover radiation and
intrinsic variations in gas-to-dust ratio. However, the range of
IRAS/DIRBE extinctions {\it around} the LMC (0.04 to 0.12) is similar
to the range, derived from the present \HI\ data (0.01 to 0.14),
implying no large systematic problem (the random errors are
negligible).  Both results are consistent with the combined Galactic
and LMC extinctions, measured by Dutra et al. (2001) using background
galaxies, of $E_{B-V}=0.12\pm0.10$.

\section{SUMMARY}

Parkes multibeam observations have been made of neutral hydrogen in
and around the Large Magellanic Cloud. The major results are:

\begin{itemize}

\item The LMC has a total \HI\ mass of $(4.8\pm0.2)\times 10^8$
  M$_{\sun}$ (for an assumed distance of 50 kpc), of which $3.8\times
  10^8$ M$_{\sun}$ lies within a well-defined disk. This is 8\% of the
  total mass and is a factor of 10 more than the molecular mass so far
  identified.

\item We measure \HI\ column densities which are 43\% higher than the
  previous survey by Luks \& Rohlfs (1992).

\item Outer tidal arms are identified. These arms appear to channel
  gas into: (a) the Magellanic Bridge, (b) the Magellanic Stream, and
  (c) the Leading Arm, and appear to be a result of the two-way
  interaction of the LMC with both the Galaxy and the Small Magellanic
  Cloud. The gas at $r>4\degr$ does not follow the outer stellar
  contours, suggesting that shocks, self-gravity or, possibly,
  external ram pressure contribute to their appearance. The existence
  of tidal shearing in the LMC argues for the presence of a massive 
  Galactic halo, and as emphasised by Weinberg (2000), suggests that 
  care needs to be taken to eliminate self-lensing when 
  LMC gravitational microlensing events are used
  to measure the density of compact objects in the Galaxy's halo.

\item High-velocity clouds, previously seen in absorption against LMC
  stars appear mainly to belong to the LMC in cases where their
  heliocentric velocity exceeds about 100 \kms, and not to the Galactic
  disk or halo. 

\item Some of the high-velocity clouds appear to be coincident with
  \HI\ voids (e.g. LMC 3 and 8), suggestive that these voids were
  created by explosive events. However, most of the \HI\ voids are
  just that -- they show no clearly defined shells of expanding gas.
  Presumably, if created by supernova and stellar winds, the giant
  voids accessible for study at the resolution of the Parkes telescope
  (i.e.  $>0.7$ kpc) have already expanded well outside of the
  thickness of the LMC disk.

\item The Galactic foreground \HI\ emission has been used to provide
  an image, at 16\farcm4 resolution, of the likely foreground dust
  extinction over the field of the LMC. The mean Galactic extinction
  is $<E_{B-V}>=0.06$ mag within the disk of the LMC. Over the full
  field imaged, there is an extinction gradient from north-west to
  south-east of $\Delta E_{B-V} \approx 0.1$ mag, corresponding to
  $\Delta A_B \approx 0.4$ mag and $\Delta A_H \approx 0.06$ mag.
\end{itemize}

\section*{Acknowledgments}

We thank Christian Br\"uns and Agris Kalnajs for helpful discussions,
the staff at the Parkes telescope for assistance during the
commissioning of the narrow-band system, and Warwick Wilson
and the AT electronics group for making the hardware and software 
work efficiently.

\label{lastpage}

\end{document}